\documentclass[reprint,superscriptaddress,preprintnumbers,nofootinbib,amsmath,amssymb,aps]{revtex4-2}

\usepackage{graphicx}
\usepackage{dcolumn}
\usepackage{bm}

\usepackage{natbib}
\usepackage{booktabs}
\usepackage{enumitem}
\usepackage{amssymb}    
\usepackage{color}         
\usepackage{graphicx}     
\usepackage{mathrsfs} 
\usepackage{hyperref} 
\hypersetup{colorlinks=true,linkcolor=blue,citecolor=blue}
\usepackage{ upgreek } 
\usepackage{color,xcolor,fancybox,epsf,rotating,colordvi}

\usepackage{ulem}

\begin{document}


\title{Jet Tagging with More-Interaction Particle Transformer}
\author{Yifan Wu}
\affiliation{College of Science, University of Shanghai for Science and Technology,  Shanghai 200093,  China}

\author{Kun Wang}
\email[]{kwang@usst.edu.cn}
\affiliation{College of Science, University of Shanghai for Science and Technology,  Shanghai 200093,  China}

\author{Congqiao Li}
\affiliation{School of Physics and State Key Laboratory of Nuclear Physics and Technology, Peking University, Beijing 100871, China}

\author{Huilin Qu}
\affiliation{CERN, EP Department, CH-1211 Geneva 23, Switzerland}

\author{Jingya Zhu}
\affiliation{School of Physics and Electronics, Henan University,  Kaifeng 475004, China}


\date{\today}

\begin{abstract}
In this study, we introduce the More-Interaction Particle Transformer (MIParT), a novel deep learning neural network designed for jet tagging. This framework incorporates our own design, the More-Interaction Attention (MIA) mechanism, which increases the dimensionality of particle interaction embeddings. We tested MIParT using the top tagging and quark-gluon datasets. Our results show that MIParT not only matches the accuracy and AUC of LorentzNet and a series of Lorentz-equivariant methods, but also significantly outperforms the ParT model in background rejection. Specifically, it improves background rejection by approximately 25\% at a 30\% signal efficiency on the top tagging dataset and by 3\% on the quark-gluon dataset. Additionally, MIParT requires only 30\% of the parameters and 53\% of the computational complexity needed by ParT, proving that high performance can be achieved with reduced model complexity. For very large datasets, we double the dimension of particle embeddings, referring to this variant as MIParT-Large (MIParT-L). We find that MIParT-L can further capitalize on the knowledge from large datasets. From a model pre-trained on the 100M JetClass dataset, the background rejection performance of the fine-tuned MIParT-L improved by 39\% on the top tagging dataset and by 6\% on the quark-gluon dataset, surpassing that of the fine-tuned ParT. Specifically, the background rejection of fine-tuned MIParT-L improved by an additional 2\% compared to the fine-tuned ParT. The results suggest that MIParT has the potential to advance efficiency benchmarks for jet tagging and event identification in particle physics.
\end{abstract}

\maketitle
\newpage


\section{\label{sec:Introduction}Introduction}

Jet identification has become a key area where machine learning is applied in high-energy physics, and has made significant progress in the past few years \cite{Larkoski:2017jix,Feickert:2021ajf}. 
Jets are collimated sprays of particles produced in high-energy collisions, typically from quarks, gluons, or the hadronic decay of heavy particles. 
The process known as jet tagging, which involves identifying the particle that initiated the jet, is complex and challenging. 
This complexity arises because the initial particle evolves into a jet through multiple stages, increasing the number of particles within the jet and obscuring the characteristics of the initiating particle.

By analyzing the constituents of a jet, it is possible to determine the type of particle that initiated the jet. 
This identification is critical for revealing fundamental physical processes and discovering new particles.
Initially, jet tagging relied heavily on quantum chromodynamics (QCD) theory, which provided methodologies for distinguishing between quark and gluon jets \cite{Gallicchio:2011xq,Gallicchio:2012ez,Larkoski:2014pca,Bhattacherjee:2015psa,FerreiradeLima:2016gcz,Gras:2017jty,Frye:2017yrw}. 
With the advent of machine learning, a variety of new jet tagging methods have been introduced that utilize different machine learning models to improve the breadth and accuracy of the techniques \cite{Cogan:2014oua,Almeida:2015jua,deOliveira:2015xxd,Komiske:2016rsd,Kasieczka:2017nvn,Macaluso:2018tck}. 
Recent advances in deep learning have further refined jet tagging methods, allowing modern algorithms to effectively process large and complex datasets. 
These algorithms are adept at identifying subtle patterns that differentiate various types of jets, significantly improving the accuracy and efficiency of jet tagging \cite{Kasieczka:2019dbj,Kagan:2020yrm,deLima:2021fwm,Kheddar:2024osf,Mondal:2024nsa}. 
The exceptional ability of deep learning to handle large data sets has been instrumental in these advances, leading to the discovery of new physical phenomena and deepening our understanding of particle interactions.

Jet tagging has undergone many changes over the years. 
Initially, traditional methods relied heavily on expert-designed features based on physical principles.
The introduction of machine learning brought more advanced approaches, starting with the concept of jet images. 
These images, representing pixelated depictions of the energy deposited by particles in a detector, marked a pivotal development in the field. 
The earliest application of jet images dates back to 1991, when Pumplin introduced the idea of representing jets as images \cite{Pumplin:1991kc}. 
Subsequent studies, starting around 2014, were inspired by computer vision. 
These studies used techniques such as Fisher's Linear Discriminant, originally used in face recognition technology, to improve jet tagging \cite{Cogan:2014oua}. 
By 2015, deep neural networks (DNNs) were being applied to top tagging \cite{Almeida:2015jua}, and later convolutional neural networks (CNNs) were widely adopted in jet tagging \cite{deOliveira:2015xxd,Komiske:2016rsd,Kasieczka:2017nvn,Lin:2018cin,Macaluso:2018tck}, demonstrating significant improvements in jet tagging performance.

In 2016, sequence-based representations began to gain traction in the field of jet tagging, using recurrent neural networks (RNNs) to process ordered data. 
This period marked a significant advancement with the pioneering use of Long Short-Term Memory (LSTM) networks for classification purposes \cite{Guest:2016iqz}. 
Subsequently, Gated Recurrent Units (GRUs) were also used for event topology classification, further extending the applications of RNNs in this domain \cite{deLima:2021fwm}.
At the same time, an innovative approach combining CNNs and LSTMs, known as DeepJet, was developed. This hybrid model significantly improved the performance of quark-gluon tagging \cite{Bols:2020bkb}. 
Additionally, several studies using RNNs introduced new methods and insights \cite{Louppe:2017ipp,Cheng:2017rdo}. 
These methods have successfully overcome the limitations associated with input size in jet tagging, providing a more flexible approach to analyzing and utilizing jet data.

In 2017, the introduction of graph-based representations using graph neural networks (GNNs) marked a significant leap forward in jet tagging \cite{Henrion:DLPS2017}. 
Subsequently, GNNs began to be widely used in particle identification, greatly expanding the capabilities of the field \cite{Abdughani:2018wrw,Martinez:2018fwc,Ren:2019xhp,Ju:2020xty}. 
This broad application of GNNs has opened new avenues for accurately classifying and understanding complex particle interactions.

In 2018, the exploration of point cloud representations, which treat jets as unordered sets of particles, marked a notable advancement. 
Komiske \textit{et al.} introduced the concept of Energy Flow Networks (EFNs), which can deal with variable-length unordered particle sets effectively \cite{Komiske:2018cqr}. 
This method utilizes the ``Deep Sets'' concept, developed by Zaheer \textit{et al.} in 2017 \cite{Zaheer:2017wmg}, which treats jets specifically as sets of particles and represents a significant advance in jet tagging. 
Crucially, it made the algorithms permutation-invariant, thereby enhancing their capability to represent complex particle interactions.

In 2019, Qu \textit{et al.} introduced ParticleNet \cite{Qu:2019gqs}, building on the Dynamic Graph Convolutional Neural Network (DGCNN) framework developed by Wang \textit{et al.} in 2018 \cite{Wang:2018nkf} . 
ParticleNet, which also treats jets as unordered sets of particles, marked significant advancements in this field. 
Recently, in 2022, Qu \textit{et al.} further extended their contributions by developing the Particle Transformer (ParT) \cite{Qu:2022mxj}, which is based on the Transformer architecture \cite{Vaswani:2017lxt}. 
By incorporating pairwise particle interaction inputs, it significantly improved performance on jet tagging.
Furthermore, the introduction of a new large-scale dataset, JetClass, enables pre-training of the ParT model, which reaches even higher performance.

However, the currently most efficient jet tagging models, the pre-trained ParT models, not only require pre-training, but also have a significant number of parameters. 
In addition, other transformer-based jet taggers fail to surpass the DGCNN-based ParticleNet due to an insufficient number of jets in the training samples. 
This indicates that transformer-based models are effective at utilizing larger training datasets by utilizing the attention mechanism. 
And we observed that pairwise particle interaction inputs play a crucial role in ParT. 
Therefore, we aim to construct a transformer-based jet tagging model with an increased focus on particle interaction inputs, aiming for optimal results without pre-training.

In this paper, we propose a new jet tagging method based on the Transformer architecture, called More-Interaction Particle Transformer (MIParT). 
We enhanced the algorithm of ParT by modifying the attention mechanism and increasing the embedding dimensions of the pairwise particle interaction inputs while reducing the total number of parameters and computational complexity. 
We tested MIParT on two widely used jet tagging benchmarks and found that it achieves improvements over existing methods. 
Additionally, to address the challenges posed by very large datasets, we doubled the particle embedding dimensions to construct a larger model. We pre-trained this enhanced model on the 100M JetClass dataset before fine-tuning it on smaller datasets. This approach showed measurable performance gains over the fine-tuned ParT, indicating the efficacy of our modifications.

The remainder of this manuscript is organized as follows.
In Sec.~\ref{sec:model}, we provide an overview of various deep learning models and specifically focus on the architecture of the MIParT.
In Sec.~\ref{sec:exp}, we detail the experimental process and follow this with an extensive discussion of the results obtained from our analysis.
In Sec.~\ref{sec:con}, we end the paper by summarizing the main conclusions and discussing their implications for future research in this area.

\section{MIParT Model Architecture}
\label{sec:model}

Traditional deep learning models such as CNNs and RNNs face significant challenges in representing jets effectively. 
Image representations often struggle with incorporating particle identity, which affects performance improvement \cite{Cogan:2014oua}. 
Similarly, sequence \cite{Guest:2016iqz} and tree \cite{Louppe:2017ipp} representations impose artificial ordering on jet particles, which inherently possess no sequential structure.
Considering a jet as an unordered collection of its constituent particles provides a more natural representation. 
This format not only facilitates the inclusion of particle-specific features, but also guarantees permutation invariance. 
Among models that adopt this perspective, ParticleNet describes jets as ``particle clouds'' drawing a parallel to the point cloud technique in 3D shape analysis in computer vision.
ParticleNet uses the DGCNN architecture, with its EdgeConv operations effectively using the local spatial structures of particle clouds to achieve significant performance improvements.

ParT, a transformative variant based on the Class-Attention in Image Transformers (CaiT) framework \cite{cord2021going}, integrates interaction variables as a secondary input.
The self-attention mechanism of this architecture uniquely addresses all positions within the input sequence, capturing extensive range dependencies efficiently and maintaining invariance to particle order. 
By refining the Multi-Head Attention (MHA) mechanism to include jet particle interaction variables, ParT not only surpasses traditional transformer models, but also sets a new benchmark in jet tagging.
These modifications position ParT as the leading model in jet tagging.

\begin{figure*}[!htbp]
\centering
\includegraphics[width=0.85\textwidth]{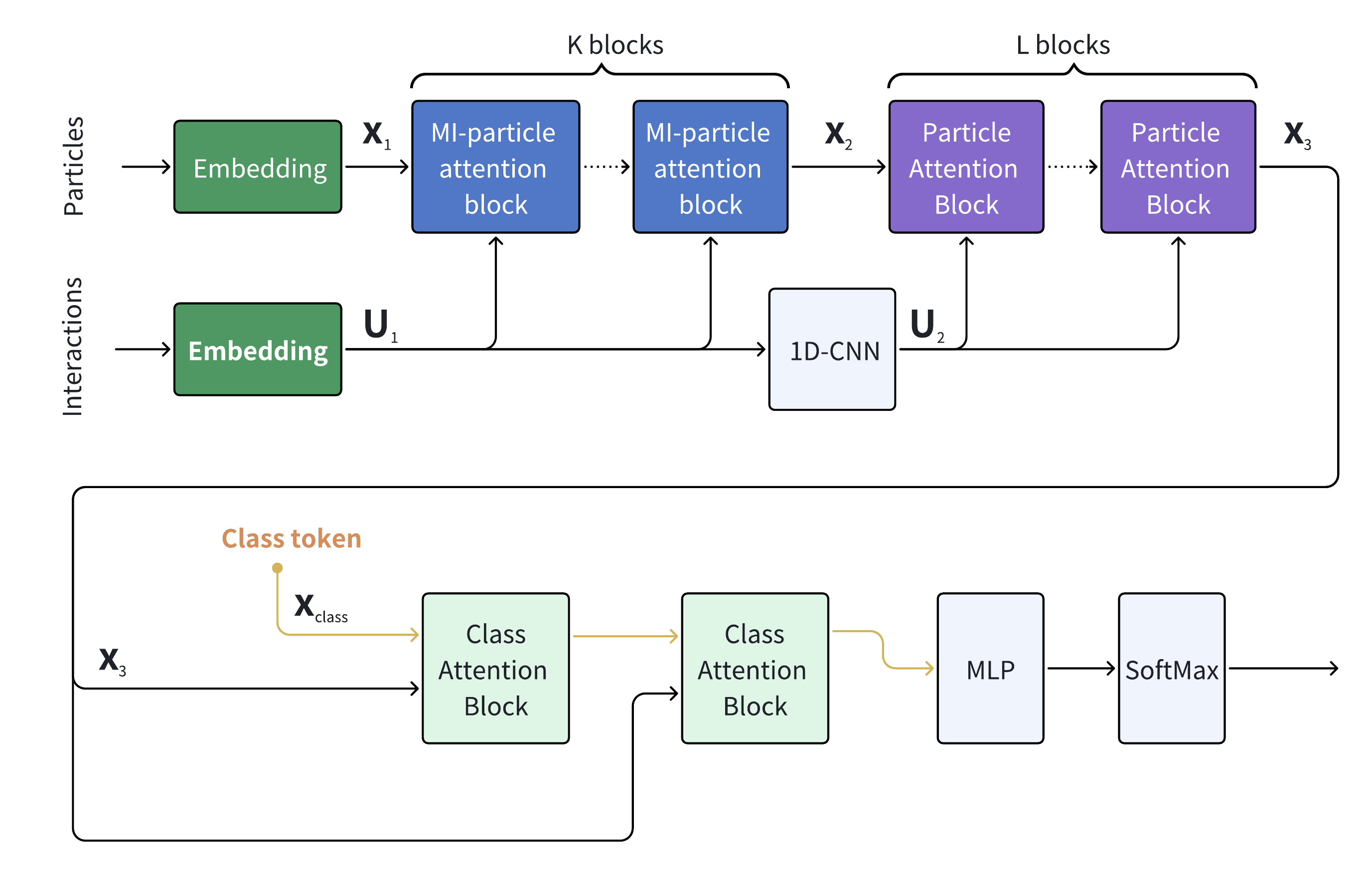}
\caption{
Schematic of the More-Interaction Particle Transformer (MIParT) architecture.
The particle features $\mathbf{x}_1$ are processed sequentially through $K$ MI particle attention blocks and $L$ particle attention blocks. The interaction features $\mathbf{U}_1$ are first fed to $K$ MI particle attention blocks, then dimensionally reduced by a 1D pointwise convolution to $\mathbf{U}_2$, and then fed to $L$ particle attention blocks.
The MIParT architecture ends with the application of the Class-Attention in Image Transformers (CaiT) methodology, which uses a class token $\mathbf{x}_{\rm class}$ to systematically extract and summarize information from $\mathbf{x}_3$ in the class attention blocks.
}
\label{fig:1}
\end{figure*}

\begin{figure}[!htbp]
\centering
\includegraphics[width=0.48\textwidth]{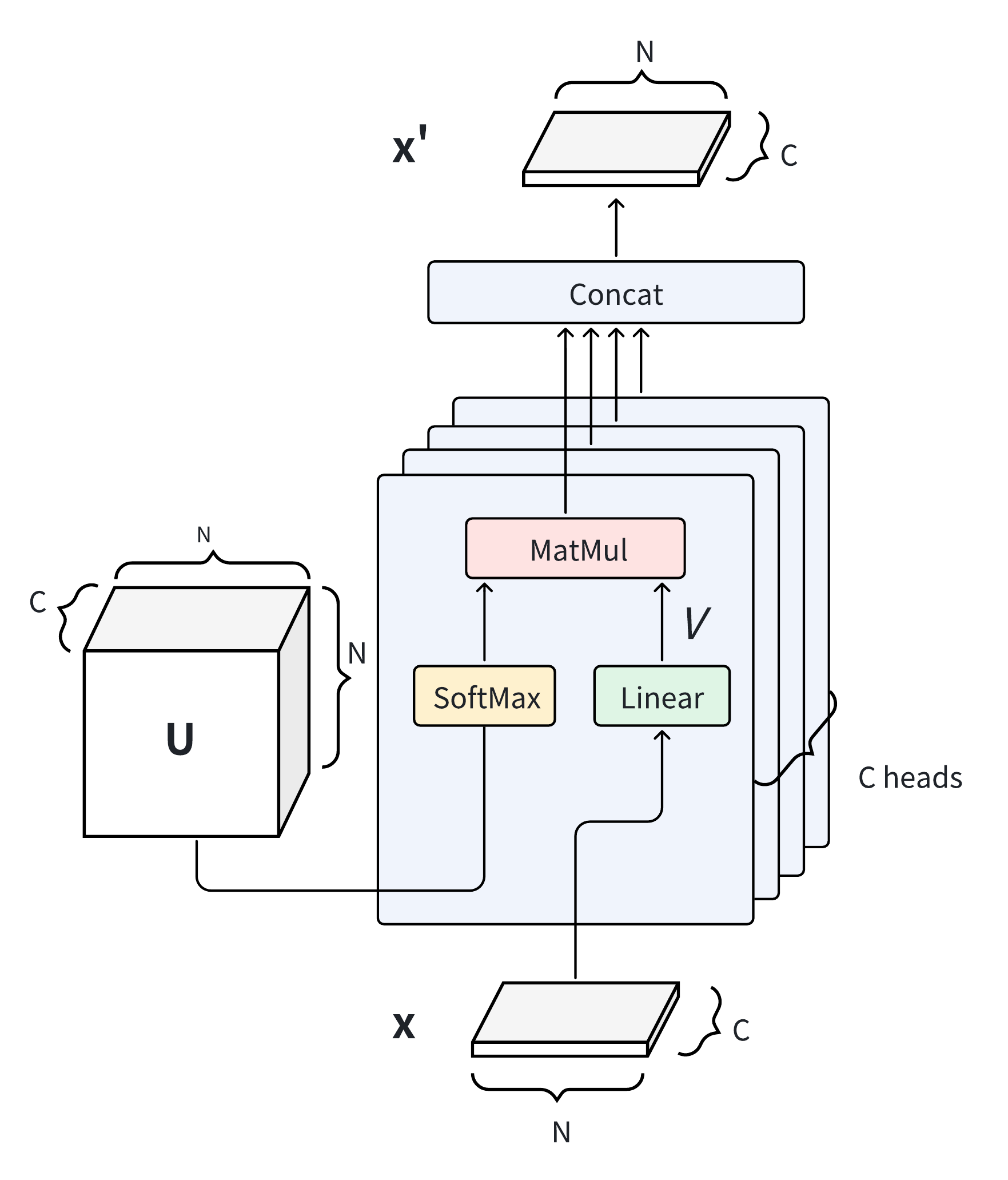}
\vspace{0.68cm}
\caption{
Schematic of the More-Interaction Attention (MIA) architecture. The shape of $\mathbf{U}$ is $(N, N, C)$, while both the input $\mathbf{x}$ and the output $\mathbf{x'}$ have the shape $(N, C)$. MIA maintains a one-to-one correspondence between the feature dimensions of $\mathbf{U}$, $\mathbf{x}$, and the heads of MHA $C$.
}
\label{fig:2}
\end{figure}

\begin{figure}[!htbp]
\centering
\includegraphics[width=0.23\textwidth]{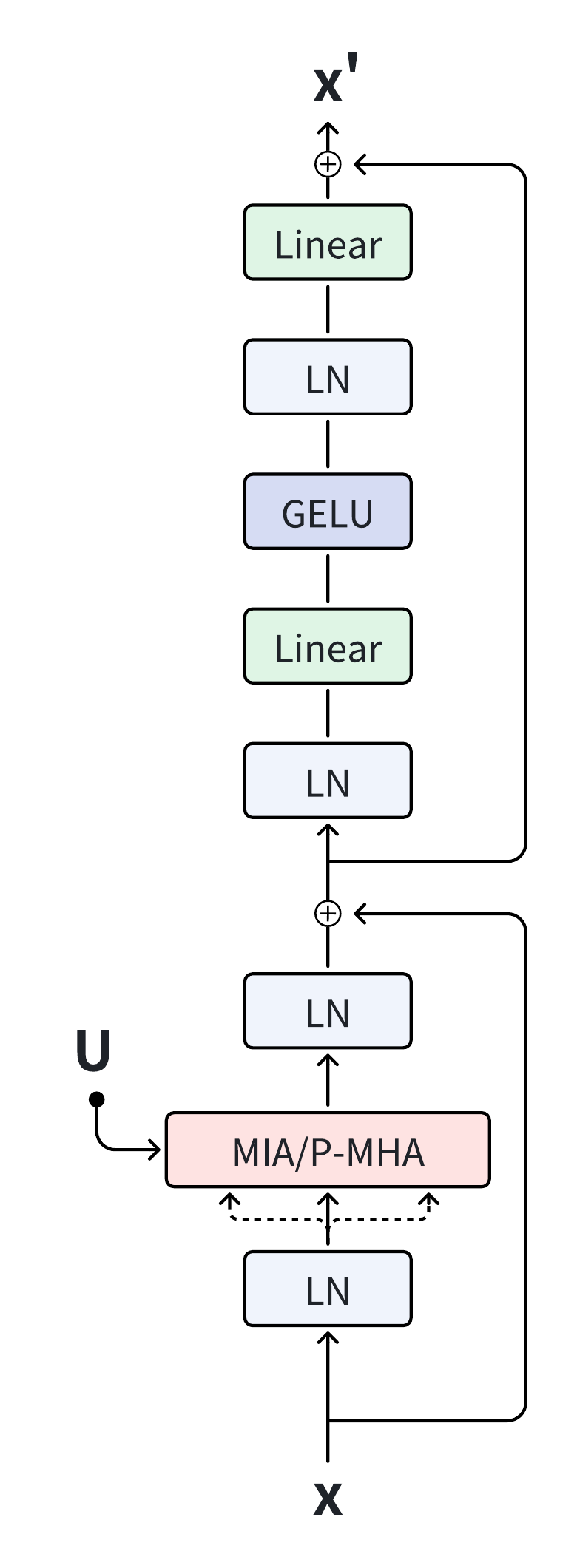}
\vspace{-0.1cm}
\caption{
Schematic of the MI-Particle Attention Block / Particle Attention Block architecture. 
Here, LN represents Layer Normalization, and GELU represents the Gaussian Error Linear Unit activation function. The block forms the MI-Particle Attention Block when using MIA and the Particle Attention Block when using P-MHA.
}
\label{fig:3}
\end{figure}

Based on the ParT framework, we develop the MIParT to enhance the input of interaction data, as depicted in Fig.~\ref{fig:1}. 
MIParT adopts ParT's input formats and processes jet data with two distinct inputs: 
\begin{itemize}
    \item Particle Input $\mathbf{x}_1$: This comprises a list of $C$ features per particle, arranged into an array of shape $(N,C)$, where $N$ represents the number of particles within a jet.
    
    \item Interaction Input $\mathbf{U}_1$: It includes a matrix of $C'$ features for each particle pair, formatted as an array of shape $(N,N,C')$.
\end{itemize}

The particle input is first transformed by a Multilayer Perceptron (MLP) to project feature dimensions to $D_1$, resulting in an array $\mathbf{x}_1$ with dimensions $(N,D_{1})$. 
Similarly, the interaction input undergoes Pointwise 1D Convolution processing, yielding $\mathbf{U}_1$ with dimensions $(N,N,D_1)$. $\mathbf{x}_1$ then passes through $K$ MI-Particle Attention Blocks to generate $\mathbf{x}_2$ of the same shape. 
In each block, $\mathbf{U}_1$ serves as an additional input and is dimensionally reduced by a Pointwise 1D Convolution to $\mathbf{U}_2$, having dimensions $(N,N,D_2)$.

Following the structural framework of ParT, $\mathbf{x}_2$ progresses through $L$ Particle Attention Blocks, enhancing with $\mathbf{U}_2$ at each layer, to produce $\mathbf{x}_3$. 
Subsequently, using the CaiT methodology, a class token $\mathbf{x}_\text{class}$ is used to systematically extract and summarize information from $\mathbf{x}_3$ in the class attention blocks. 
Finally, this summarized information forms a single vector that is input into a linear classifier through an MLP and a softmax function to derive the classification scores.

\subsection{Particle Attention Block}

The Particle Attention Block, a crucial element of the ParT framework, has been seamlessly integrated into our MIParT model. 
The architecture of this block is based on the NormFormer design \cite{shleifer2021normformer}, specifically using the Layer Normalization instead of the Batch Normalization. 
Layer Normalization optimizes normalization by adjusting each layer individually for every single sample, enhancing model stability and overall performance across diverse datasets. 
The architecture of the Particle Attention Block is illustrated in Fig.~\ref{fig:3}.
Furthermore, in this configuration, the traditional Multi-Head Attention (MHA) is substituted by Particle-Multi-Head Attention (P-MHA). 
This modification allows for the incorporation of particle interaction features directly into the attention mechanism, enriching the model's capability to capture complex particle dynamics.
The P-MHA mechanism, which is key to the Particle Attention Block, is mathematically expressed as
\begin{align}
    \text{P-MHA}(Q, K, V) = \text{SoftMax}\left(\frac{QK^T}{\sqrt{d_k}} + \mathbf{U}\right)V,
\end{align}
where $Q$, $K$, and $V$ are the linear projections of the particle embedding $x$, and $\mathbf{U}$ represents the interaction embedding. 
The dimensions of $\mathbf{U}$ are precisely aligned with the attention heads in the MHA mechanism, thereby facilitating the integration of particle interaction features. 
The specific implementation of P-MHA can be found in Ref.~\cite{Qu:2022mxj}. 
This integration significantly enhances the model's ability to capture complex particle interactions, which is crucial in particle physics applications.

\subsection{MI-Particle Attention Blocks}

In the original P-MHA mechanism, the feature dimensions of $\mathbf{U}$ align one-to-one with the heads of MHA, both denoted as $C$. 
Increasing the feature dimensions of $\mathbf{U}$ necessitates a proportional increase in the number of attention heads, which significantly adds to the model's complexity. 
To mitigate this issue, we introduce More-Interaction Attention (MIA) and the MI-Particle Attention Block. 
These components replace P-MHA with MIA, as illustrated in Fig.~\ref{fig:2} (MIA architecture) and Fig.~\ref{fig:3} (MI-Particle Attention Block/Particle Attention Block architecture). 
The MI-Particle Attention Block incorporates Layer Normalization and the Gaussian Error Linear Unit (GELU) activation function. When the red block in the Fig.~\ref{fig:3} uses MIA, it forms the MI-Particle Attention Block. Conversely, when it uses P-MHA, it forms the Particle Attention Block. 
This approach allows the model to effectively use interaction inputs without significantly increasing complexity.
The MIA is calculated using the following formula:
\begin{align}
    \text{MIA}(\mathbf{U}, V) = \text{SoftMax}(\mathbf{U})V,
\end{align}
where $V$ is a linear projection of the particle embedding $\mathbf{x}$. In MIA, each feature dimension of $\mathbf{U}$ and $\mathbf{x}$, as well as each head, are denoted by $C$, ensuring a one-to-one correspondence.

By increasing the feature dimensions of $\mathbf{U}$, MIA effectively exploits the interaction inputs without significantly increasing the complexity of the model. 
Moreover, the MI-Particle Attention Block, which incorporates self-attention on $\mathbf{x}$, acts as a supplement in front of the Particle Attention Block rather than replacing it.

\begin{figure}[!htbp]
\centering
\includegraphics[width=0.24\textwidth]{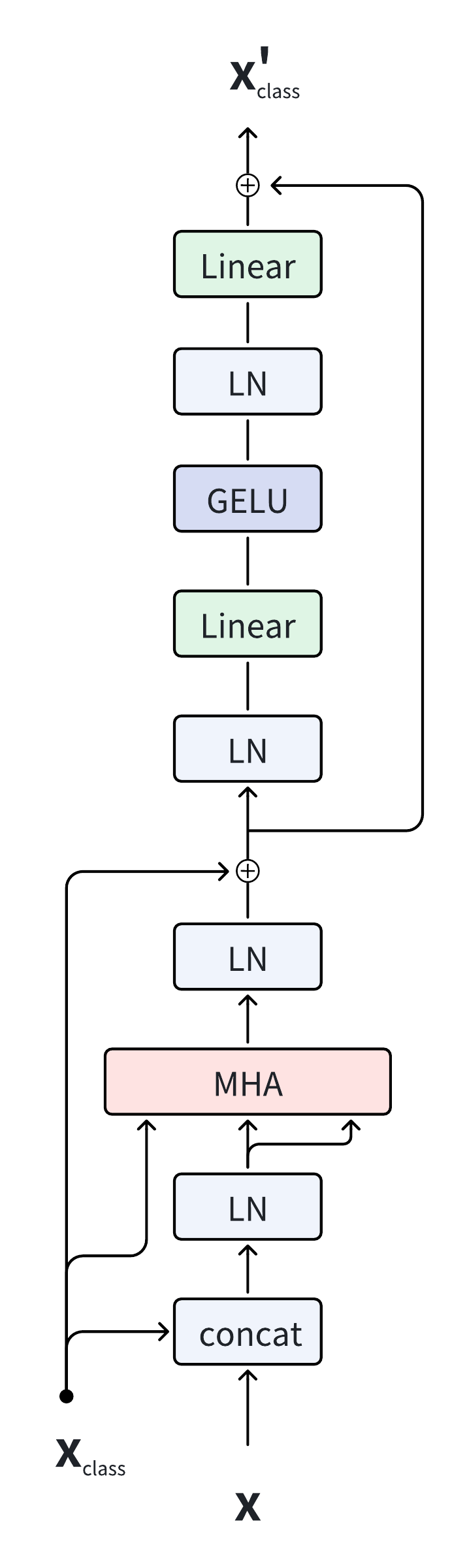}
\caption{
Schematic of the Class Attention Block architecture. 
Here, LN represents Layer Normalization, GELU represents the Gaussian Error Linear Unit activation function, and MHA stands for the Multi-Head Attention block. 
}
\label{fig:4}
\end{figure}

\subsection{Class Attention Block}

We incorporated the Class Attention Block from the ParT framework, inspired by the CaiT architecture. 
This block uses a class token $\mathbf{x}_\text{class}$ to efficiently extract information through attention mechanisms, as depicted in Fig.~\ref{fig:4}. 
The Multi-Head Attention inputs are defined as follows:
\begin{align}
    Q &= W_{q}\mathbf{x}_\text{class}+b_{q},\\
    K &= W_{k}\mathbf{z}+b_{k},\\
    V &= W_{v}\mathbf{z}+b_{v}
\end{align}
where $\mathbf{z}=[\mathbf{x}_\text{class},\mathbf{x}]$, and $W$ and $b$ represent learnable parameters. 
This design ensures a lower computational overhead for the Class Attention mechanism by utilizing the concatenated vector $\mathbf{z}$. 

The Class Attention Block significantly enhances feature extraction from the input $\mathbf{x}$ by capitalizing on the class token, thereby improving the model’s focus on essential aspects of the data. 
This enhancement significantly improves jet classification performance, making the Class Attention Block as a crucial component within the ParT framework.

\subsection{Implementation}
The architecture of our MIParT model includes $K = 5$ MI-particle attention blocks, $L = 5$ particle attention blocks, and 2 class attention blocks. 
The choice of these hyperparameters balances complexity and accuracy; we observed an increase in accuracy with additional layers, but at the cost of increased complexity. Therefore, we limited the total number of attention blocks to ten. 
The rationale for choosing two class attention blocks follows the CaiT framework \cite{cord2021going}, which recommends such a configuration for efficient classification.
For particle embeddings $\mathbf{x}_1$, a three-layer Multi-Layer Perceptron (MLP) is used, with each layer containing 128, 512, and 64 neurons respectively.
This configuration results in embeddings with a dimensionality of $D_1 = 64$. 
The decision to reduce the embedding dimension compared to the ParT model was motivated by the addition of the MIA module. This adjustment allows us to rationalize the complexity of the model while maintaining its efficiency, thus optimizing the trade-off between performance and computational load.
Each layer incorporates GELU as the activation function and Layer Normalization.
Additionally, a three-layer, 64-channel Pointwise 1D Convolution is used for the interaction embeddings $\mathbf{U}_1$, performing convolutions only along the feature dimension. 
The $\mathbf{U}_1$ embeddings are further processed through a single-layer, 8-channel Pointwise 1D Convolution to generate $\mathbf{U}_2$, achieving a dimensionality of $D_2 = 8$.
This design choice maintains consistency with the ParT model, ensuring alignment with established architectural standards and facilitating comparative analysis.
The MI-particle attention blocks implement MIA with 64 heads, while the P-MHA and Class Multi-Head Attention in the particle and class attention blocks utilize 8 heads each. 
A dropout rate of 0.1 is maintained in all MI-particle and particle attention blocks, with the class attention blocks being exempt from dropout.

For very large datasets, increasing the embedding dimension significantly enhances model performance. Therefore, for such datasets, we double the dimension of the particle embeddings to $D_1 = 128$. This adjustment is straightforward, requiring a change in the neuron configuration of the three-layer MLP to 128, 512, and 128. Consequently, the dimensions of $\mathbf{x}$ and $\mathbf{U}$ in MIA will no longer be identical; however, this discrepancy is acceptable as long as the dimension of $\mathbf{x}$ is an integer multiple of the dimension of $\mathbf{U}$. We refer to this modified model as MIParT-Large (MIParT-L).

\section{Result and Discussion}
\label{sec:exp}

\begin{table}[!tbp]
\centering
\caption{\label{tab:1} 
Summary of kinematic and particle identification variables included in the top tagging (TOP), quark-gluon (QG) and JetClass (JC) datasets. Variables present in each dataset are indicated by a star symbol ($\star$). The table includes seven kinematic variables describing the physical characteristics of particles relative to the jet axis, six particle identification variables categorizing particles by type and charge, and four trajectory displacement features, which provide detailed information on particle trajectories.
}
\resizebox*{\columnwidth}{!}{
\setlength{\tabcolsep}{8.0pt}
\begin{tabular}{@{\hspace{7pt}}ccccc@{\hspace{7pt}}}
\toprule
Category &      Variable &  TOP & QG & JC \\ \midrule
         &   $\Delta\eta$  &        $\star$         &$\star$ &$\star$   \\
         &  $\Delta\phi$&        $\star$       &   $\star$ &$\star$ \\
    &  log $p_{\rm T}$    &        $\star$         &  $\star$ &$\star$ \\
   Kinematics       & log $E$ &        $\star$        &    $\star$ &$\star$ \\
         &   $\log {p_{\rm T}}/{p_{\rm T}{\rm(jet)}}$&      $\star$    &   $\star$ &$\star$ \\
         &  $\log { E}/{ E{\rm(jet)}}$&    $\star$      & $\star$  &$\star$ \\
         &  $\Delta R$ &      $\star$          & $\star$  &$\star$ \\ \midrule
         &  charge  &                 & $\star$ &$\star$ \\
         &   Electron   &                & $\star$  &$\star$ \\
  Particle   &   Muon    &                &   $\star$  &$\star$ \\
identification    &  Photon     &                &   $\star$ &$\star$ \\
         &  Charged Hadron    &                 &  $\star$  &$\star$ \\
         &    Neutral Hadron   &               & $\star$  &$\star$ \\   \midrule
         & $\tanh d_0$& & &$\star$ \\
    Trajectory     & $\tanh d_z$& & &$\star$ \\
    displacement    & $\sigma_{d_0}$& & &$\star$ \\
         & $\sigma_{d_z}$& & &$\star$ \\
         \bottomrule
\end{tabular}}
\end{table}

We developed the MIParT model using the PyTorch framework \cite{paszke2019pytorch}, implemented based on the \textsf{Weaver}\footnote{$\textsf{Weaver}$ provides a streamlined yet flexible machine learning R\&D framework for high energy physics, \url{https://github.com/hqucms/weaver-core}.}, and also referred to the implementation of ParT\footnote{The official implementation of Particle Transformer for Jet Tagging, which includes the code and pre-trained models, \url{https://github.com/jet-universe/particle_transformer}.}.

We initially evaluated the MIParT model on two widely used jet tagging benchmark datasets, top tagging \cite{Kasieczka:2019dbj} and quark-gluon datasets \cite{komiske2019energy}. 
The model was trained on an NVIDIA RTX 4090 GPU, using a learning rate of 0.001 and a batch size of 256. Training was limited to 15 epochs to prevent overfitting.
Both datasets incorporate kinematic variables as particle input features, with particle identification information included only in the quark-gluon dataset. All these input features for the two datasets are shown in Table~\ref{tab:1}.

We then pre-trained our larger model variant, MIParT-L, on the JetClass dataset containing 100M samples \cite{Qu:2022mxj}. This model was pre-trained on dual NVIDIA RTX 3090 GPUs using a learning rate of 0.0008 and a batch size of 384, with pre-training limited to 50 epochs to avoid overfitting. After pre-training, MIParT-L was fine-tuned on the top tagging and quark-gluon datasets. It is noteworthy that the pre-training of MIParT-L on the JetClass dataset for the top tagging dataset included only kinematic features, while for the quark-gluon dataset both kinematic and particle identification features were included.

For fine-tuning, we replaced the last MLP for classification with a newly initialized MLP having two output nodes. All weights were then fine-tuned across the datasets for 20 epochs. We used a learning rate of 0.00016 for the pre-trained weights and 0.008 for the new MLP.

The seven kinematic input features are:
\begin{itemize}
\item $\Delta\eta$: the difference in pseudorapidity $\eta$ between the particle and the jet axis;
\item $\Delta\phi$: the difference in azimuthal angle $\phi$ between the particle and the jet axis;
\item $\log p_{\rm T}$: the logarithm of the particle's transverse momentum $p_{\rm T}$;
\item $\log E$: the logarithm of the particle's energy;
\item $\log {p_{\rm T}}/{p_{\rm T}{\rm (jet)}}$: the logarithm of the particle's $p_{\rm T}$ relative to the jet $p_{\rm T}$;
\item $\log {E}/{E{\rm (jet)}}$: the logarithm of the particle's energy relative to the jet energy;
\item $\Delta R$: the angular separation between the particle and the jet axis.
\end{itemize}

The six particle identification features are:
\begin{itemize}
    \item ``Charge": the electric charge of the particle;
    \item ``Electron": whether the particle is an electron;
    \item ``Muon": whether the particle is a muon;
    \item ``Photon": whether the particle is a photon;
    \item ``Charged Hadron": whether the particle is a charged hadron;
    \item ``Neutral Hadron": whether the particle is a neutral hadron.
\end{itemize}

The four trajectory displacement features in the JetClass are:
\begin{itemize}
    \item $\tanh d_0$: hyperbolic tangent of the transverse impact parameter value;
    \item $\tanh d_z$: hyperbolic tangent of the longitudinal impact parameter value;
    \item $\sigma_{d_0}$: error of the measured transverse impact parameter;
    \item $\sigma_{d_z}$: error of the measured longitudinal impact parameter.
\end{itemize}

\begin{table*}[!tbp]
\centering
\caption{\label{tab:2}
Comparative performance of various models on the top tagging dataset. This table displays the results for the MIParT model alongside those of other prominent models such as Particle Flow Network (PFN) \cite{komiske2019energy}, Particle-level Convolutional Neural Network (P-CNN),  Point Cloud Transformer (PCT) \cite{Mikuni:2021pou}, Clifford Group Equivariant Neural Networks (CGENN) \cite{Ruhe:2023rqc}, Permutation equivariant and lorentz invariant or covariant aggregator network (PELICAN) \cite{Bogatskiy:2022czk}, Lorentz-Equivariant Geometric Algebra Transformers (L-GATr) \cite{Spinner:2024hjm}, LorentzNet \cite{Gong:2022lye}, ParticleNet \cite{Qu:2019gqs}, ParT \cite{Qu:2022mxj}. Metrics of other models are quoted from their published results.
The fine-tuned version of our model, MIParT-L f.t., is displayed at the bottom of the table for comparison with the fine-tuned ParT model, ParT f.t.
}
\resizebox*{1\textwidth}{!}{
\setlength{\tabcolsep}{25pt}
\begin{tabular}{@{\hspace{20pt}}cllll@{\hspace{20pt}}}
\toprule
               & Accuracy       & AUC             & $\text{Rej}_{50\%}$         & $\text{Rej}_{30\%}$           \\ \midrule
PFN            & —              & 0.9819          & 247±3           & \ 888±17            \\
P-CNN          & 0.930          & 0.9803          & 201±4           & \ 759±24            \\
PCT            & 0.940          & 0.9855          & 392±7           & 1533±101          \\
CGENN          & 0.942          &  0.9869          &  500          & 2172            \\
PELICAN        & \textbf{0.9426}         &  \textbf{0.9870}         & —          & —               \\
L-GATr         &  0.9417        &  0.9868          & \textbf{548±26}         & 2148±106     \\
LorentzNet     & 0.942          & 0.9868 & 498±18          & \textbf{2195±173} \\
ParticleNet    & 0.940          & 0.9858          & 397±7           & 1615±93           \\
ParT           & 0.940          & 0.9858          & 413±16          & 1602±81           \\
\textbf{MIParT (ours)} &   0.942 & 0.9868 & 505±8 & 2010±97      \\ 
\midrule
ParT f.t.       & \textbf{0.944}    & 0.9877 & \textbf{691±15} & 2766±130 \\
\textbf{MIParT-L f.t. (ours)}      & \textbf{0.944}    & \textbf{0.9878} & 640±10    & \textbf{2789±133}   \\ \bottomrule
\end{tabular}}
\end{table*}

\begin{table*}[!tbp]
\centering
\caption{\label{tab:3} 
Comparative performance of various models on the quark-gluon dataset. This table outlines the results for the MIParT model along with other significant models, including Particle Flow Network (PFN) \cite{komiske2019energy}, attention-based Cloud Net (ABCNet) \cite{Mikuni:2020wpr}, Point Cloud Transformer (PCT) \cite{Mikuni:2021pou}, LorentzNet \cite{Gong:2022lye}, and ParT \cite{Qu:2022mxj}. Metrics of other models are cited from their published results. 
The fine-tuned version of our model, MIParT-L f.t., is displayed at the bottom of the table for comparison with the fine-tuned ParT model, ParT f.t.
}
\resizebox*{1\textwidth}{!}{
\setlength{\tabcolsep}{25.0pt}
\begin{tabular}{@{\hspace{20pt}}ccccc@{\hspace{20pt}}}
\toprule
           & Accuracy & AUC    & $\text{Rej}_{50\%}$  & $\text{Rej}_{30\%}$   \\ \midrule
PFN        & ---        & 0.9052 & 37.4±0.7 & ---         \\
ABCNet     & 0.840    & 0.9126 & 42.6±0.4 & 118.4±1.5 \\
PCT        & 0.841    & 0.9140 & 43.2±0.7 & 118.0±2.2 \\
LorentzNet & 0.844    & 0.9156 & 42.4±0.4 & 110.2±1.3 \\
ParT       & 0.849    & 0.9203 & 47.9±0.5 & 129.5±0.9 \\
\textbf{MIParT (ours)}      & \textbf{0.851}    & \textbf{0.9215} & \textbf{49.3±0.4}    & \textbf{133.9±1.4}   \\ \midrule
ParT f.t.       & 0.852    & 0.9230 & 50.6±0.2 & 138.7±1.3 \\
\textbf{MIParT-L f.t. (ours)}      & \textbf{0.853}    & \textbf{0.9237} & \textbf{51.9±0.5}    & \textbf{141.4±1.5}   \\

\bottomrule
\end{tabular}}
\end{table*}

For particle interaction features, we consider four logarithmic characteristics $(\ln \Delta, \ln k_T, \ln z, \ln m^2)$ derived from the energy-momentum four-vector $p = (E, p_x, p_y, p_z)$ \cite{Dreyer:2020brq}. 
These features are defined as follows: 
\begin{align}
    \Delta&= \sqrt{(y_a -y_b)^2 +(\phi_a -\phi_b)^2},\\
      k_T &= {\rm min}(p_{{\rm T},a},p_{{\rm T},b})\Delta,\\
      z &={\rm min}(p_{{\rm T},a},p_{{\rm T},b})/(p_{{\rm T},a} + p_{{\rm T},b}),\\
      m^2 &=(E_a+E_b)^2 - |{\mathbf{p}_a+\mathbf{p}_b}|^2 \, ,
\end{align}
where $y_i$ is the rapidity, $\phi_i$ is the azimuthal angle, $p_{{\rm T},i}$ is the transverse momentum, and $\mathbf{p}_i$ is the momentum 3-vector of the particle $i=a,b$.
The motivation for selecting these variables comes from their widespread adoption in several advanced neural networks \cite{Qu:2019gqs, Qu:2022mxj}.

\begin{figure*}[!htbp]
\centering
\includegraphics[width=0.72\textwidth]{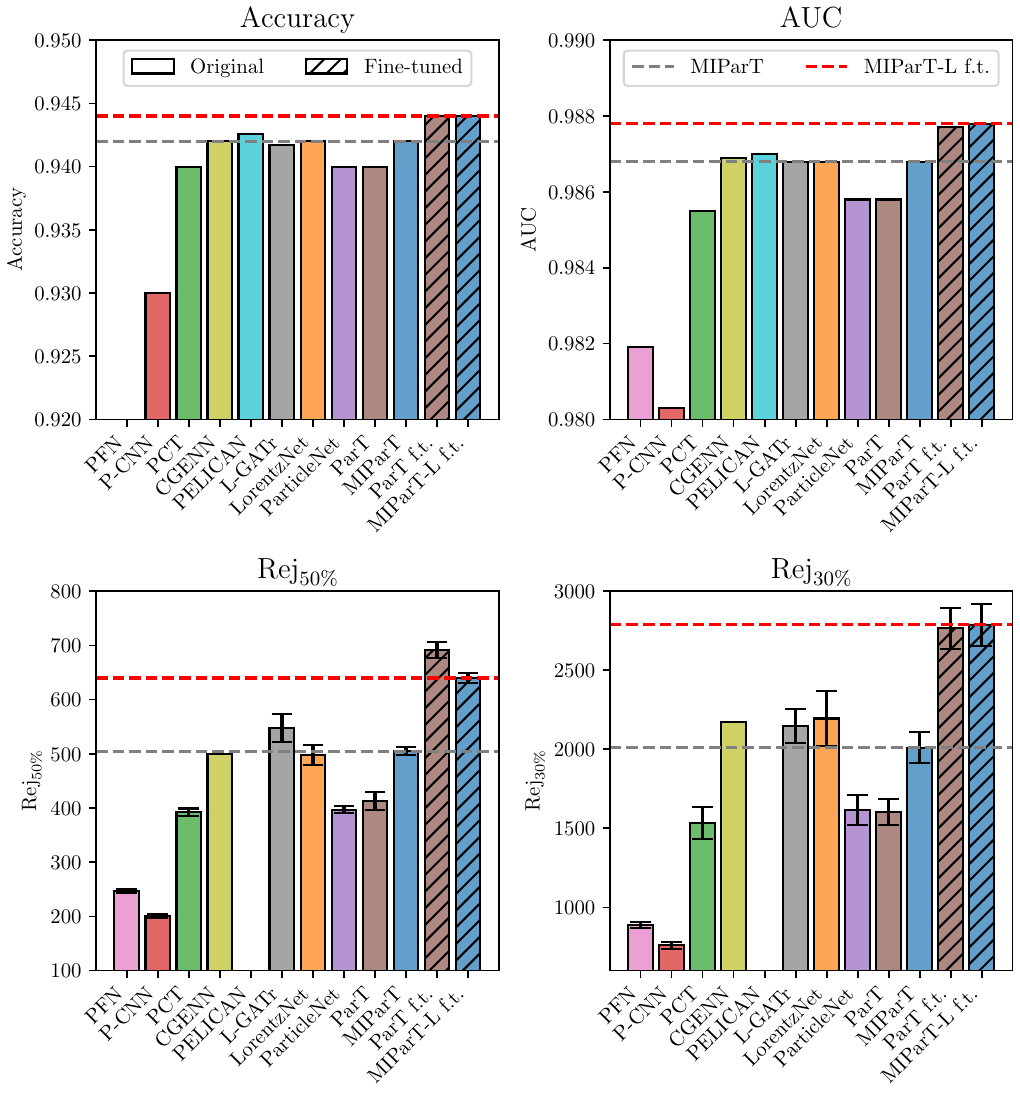}
\caption{
Performance metrics comparison of MIParT with other models on the top tagging dataset. This figure displays the Accuracy, AUC, ${\rm Rej}_{50\%}$, and ${\rm Rej}_{30\%}$ metrics for the MIParT model alongside Particle Flow Network (PFN) \cite{komiske2019energy}, Particle-level Convolutional Neural Network (P-CNN),  Point Cloud Transformer (PCT) \cite{Mikuni:2021pou}, Clifford Group Equivariant Neural Networks (CGENN) \cite{Ruhe:2023rqc}, Permutation equivariant and lorentz invariant or covariant aggregator network (PELICAN) \cite{Bogatskiy:2022czk}, Lorentz-Equivariant Geometric Algebra Transformers (L-GATr) \cite{Spinner:2024hjm}, LorentzNet \cite{Gong:2022lye}, ParticleNet \cite{Qu:2019gqs}, ParT \cite{Qu:2022mxj}. Metrics of other models are quoted from their published results. Detailed outcomes are provided in Table~\ref{tab:2}. 
Bars without slashes indicate the original models without fine-tuning, while bars with slashes indicate models with fine-tuning. The gray dashed line indicates the results for MIParT, and a red dashed line shows the results for the fine-tuned MIParT-L (MIParT-L f.t.).
}
\label{fig:5}
\end{figure*}

\begin{figure*}[!htbp]
\centering
\includegraphics[width=0.72\textwidth]{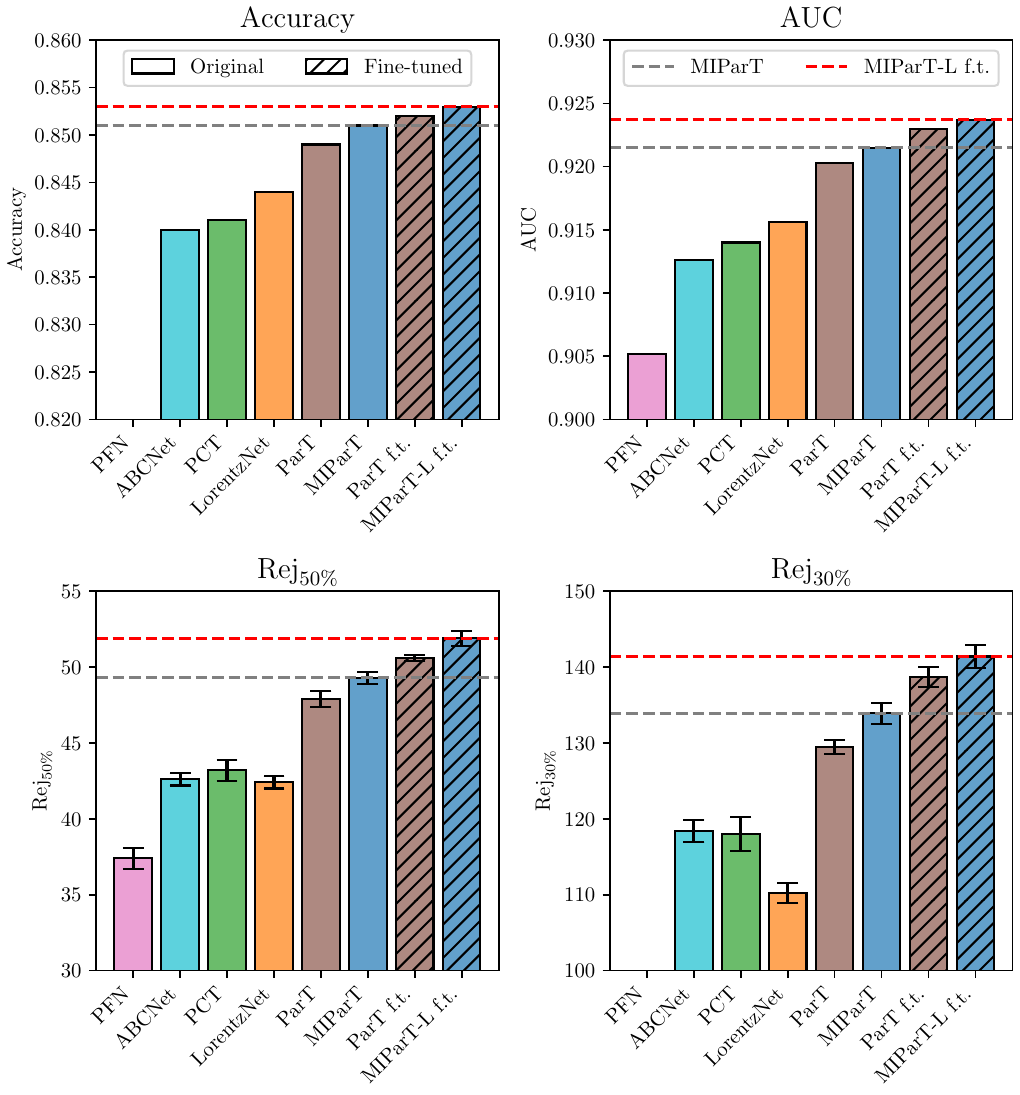}
\caption{
Performance metrics comparison of MIParT with other models on the quark-gluon dataset. This figure displays the Accuracy, AUC, ${\rm Rej}_{50\%}$, and ${\rm Rej}_{30\%}$ metrics for the MIParT model alongside Particle Flow Network (PFN) \cite{komiske2019energy}, attention-based Cloud Net (ABCNet) \cite{Mikuni:2020wpr}, Point Cloud Transformer (PCT) \cite{Mikuni:2021pou}, LorentzNet \cite{Gong:2022lye}, and ParT \cite{Qu:2022mxj}. Metrics of other models are quoted from their published results. Detailed outcomes are provided in Table~\ref{tab:2}. 
Bars without slashes indicate the original models without fine-tuning, while bars with slashes indicate models with fine-tuning. The gray dashed line indicates the results for MIParT, and a red dashed line shows the results for the fine-tuned MIParT-L (MIParT-L f.t.).
}
\label{fig:6}
\end{figure*}

To evaluate the performance of the MIParT model, we conducted comparative evaluations with several popular models using the top tagging and quark-gluon datasets. Our evaluation focused on several key metrics:
\begin{itemize}
    \item \textbf{Accuracy}: This metric quantifies the proportion of correct predictions made by the model, including both true positive and true negative identifications. Mathematically, accuracy is defined as: 
    \begin{align}
        \text{Accuracy} = \frac{TP+TN}{TP+TN+FN+FP} \, ,
    \end{align}
    where $TP$ is true positives, $TN$ is true negatives, $FN$ is false negatives, and $FP$ is false positives.

    \item \textbf{AUC (Area Under the Curve)}: AUC provides a comprehensive measure of model performance across all classification thresholds. This metric is derived from the Receiver Operating Characteristic (ROC) curve, which plots the true positive rate (sensitivity) against the false positive rate ($1 - \text{specificity}$) for various thresholds. This curve illustrates the trade-off between sensitivity and specificity. An AUC value can range from 0.5, which indicates no discriminatory ability (similar to random guessing), to 1.0, which represents perfect discrimination and indicates the model's excellent ability to discriminate between classes.

    \item \textbf{Background Rejection at a Certain Signal Efficiency, ${\rm Rej}_{X\%}$}: This metric calculates the inverse of the false positive rate (FPR) when the true positive rate (TPR) is fixed at a certain percentage, commonly referred to as ${\rm Rej}_{X\%}$. It is mathematically expressed as:
    \begin{align}
    {\rm Rej}_{X\%} = \frac{1}{\text{FPR}}\bigg|_{\text{TPR} = X\%}
    \end{align}
    For example, a ${\rm Rej}_{30\%}$ value of 2500 indicates that at a TPR of 30\%, the inverse of the FPR is 2500. This equates to only one false positive for every 2500 negative instances, highlighting the exceptional specificity and minimal error rate of the model at this level.     
\end{itemize}

Top tagging is a critical task in jet tagging, which is often used in the search for new physics at the LHC. 
For this study, we used a top tagging dataset \cite{Kasieczka:2019dbj} consisting of 2M jets, with $t \to bqq'$ as the signal and $q/g$ as the background. 
This dataset only provides the energy-momentum four-vectors (kinematic features) for each particle.

In Fig.~\ref{fig:5}, we showed the performance of our MIParT model compared to other popular models on the top tagging dataset. 
The MIParT model achieved accuracy and AUC metrics nearly identical to those of LorentzNet \cite{Gong:2022lye}, and its ${\rm Rej}_{50\%}$ and ${\rm Rej}_{30\%}$ metrics are within the error range comparable to LorentzNet. 
It is noteworthy that a series of Lorentz-equivariant methods demonstrated performance similar to that of LorentzNet, such as Clifford Group Equivariant Neural Networks (CGENN) \cite{Ruhe:2023rqc}, Permutation equivariant and Lorentz invariant or covariant aggregator network (PELICAN) \cite{Bogatskiy:2022czk}, Lorentz-Equivariant Geometric Algebra Transformers (L-GATr) \cite{Spinner:2024hjm}. 
Moreover, MIParT, LorentzNet, and several Lorentz-equivariant based models significantly outperformed other models, including Particle Flow Network (PFN) \cite{komiske2019energy}, Particle-level Convolutional Neural Network (P-CNN) \cite{Qu:2019gqs}, ParticleNet \cite{Qu:2019gqs}, Point Cloud Transformer (PCT) \cite{Mikuni:2021pou}, and ParT \cite{Qu:2022mxj}, with metrics quoted from their published results. 
For the fine-tuned MIParT-L model that is pre-trained on the 100M JetClass dataset, a 39\% enhancement in background rejection performance was achieved, comparable to that of the fine-tuned ParT. 
Detailed comparison results are presented in Table~\ref{tab:2}. 
The MIParT model significantly outperformed ParT in the top tagging benchmark, with  approximately 25\% better background rejection at a 30\% signal efficiency. Among the models evaluated,  MIParT, along with LorentzNet and several other Lorentz-equivariant based models, ranks at the top tier, showcasing some of the most robust performances.

\begin{table*}[!tbp]
\centering
\caption{\label{tab:flop}
Parameters, FLOPs, and Accuracy of various models on the top tagging (TOP) and quark-gluon (QG) datasets. Parameters refer to the number of trainable elements within a model, while FLOPs (Floating Point Operations Per Second) measure the computational complexity involved in processing data through the model.
}
\resizebox*{1\textwidth}{!}{
\setlength{\tabcolsep}{30.0pt}
\begin{tabular}{@{\hspace{20pt}}ccccc@{\hspace{20pt}}}
\toprule
            & TOP   & QG   & Params & FLOPs \\ \midrule
PFN         & ---     &    & 86.1k  & 4.62M \\
P-CNN       & 0.930 & ---     & 354k   & 15.5M \\
ParticleNet & 0.940 & ---     & 370k   & 540M  \\
ParT        & 0.940 & 0.849 & 2.14M  & 340M  \\
\textbf{MIParT (ours)}      & \textbf{0.942} & \textbf{0.851} & 720.9k & 180M  \\ 
\textbf{MIParT-L f.t. (ours)}      & \textbf{0.944} & \textbf{0.853} & 2.38M & 368M  \\ 
\bottomrule
\end{tabular}}
\end{table*}

\begin{table*}[tb]
\centering
\caption{Comparative performance of various models on different sizes of the JetClass dataset. This table outlines the results for the MIParT-L model alongside ParticleNet \cite{Qu:2019gqs} and ParT \cite{Qu:2022mxj} across 2M, 10M, and 100M JetClass datasets. Metrics of other models are cited from their published results. Models trained using the full 100M training dataset are highlighted in bold text.}
\label{tab:jc}
\resizebox*{1\textwidth}{!}{
\setlength{\tabcolsep}{1.3pt}
\begin{tabular}{cccccccccccc}
\toprule
    & \multicolumn{2}{c}{All classes}   & $H\to b \bar{b}$      & $H\to c \bar{c}$     & $H\to g g $      & $H\to 4 q$    & $H\to \ell \nu q q'$    & $t\to b q q'$     & $t\to b \ell \nu$     & $W\to q q'$      & $Z\to q q'$ \\
    & Accuracy  & AUC       &$\text{Rej}_{50\%}$  &$\text{Rej}_{50\%}$ &$\text{Rej}_{50\%}$ &$\text{Rej}_{50\%}$ &$\text{Rej}_{99\%}$ &$\text{Rej}_{50\%}$&$\text{Rej}_{99.5\%}$&$\text{Rej}_{50\%}$ &$\text{Rej}_{50\%}$ \\
\midrule
ParticleNet (2\,M)                   & 0.828     & 0.9820    & 5540      & 1681      & 90        & 662       & 1654      & 4049      & 4673      & 260       & 215       \\
ParticleNet (10\,M)                  & 0.837     & 0.9837    & 5848      & 2070      & 96       & 770      & 2350      & 5495     & 6803      & 307       & 253       \\
\bf ParticleNet (100\,M)             & 0.844     & 0.9849    & 7634     & 2475      & 104       & 954      & 3339      & 10526     & 11173     & 347       & 283       \\
\midrule
ParT (2\,M)                   & 0.836     & 0.9834    & 5587      & 1982      & 93        & 761       & 1609      & 6061      & 4474      & 307       & 236       \\
ParT (10\,M)                  & 0.850     & 0.9860    & 8734      & 3040      & 110       & 1274      & 3257      & 12579     & 8969      & 431       & 324       \\
\bf ParT (100\,M)             & 0.861     & 0.9877    & 10638     & 4149      & 123       & 1864      & 5479      & 32787     & 15873     & 543       & 402       \\
\midrule
MIParT-L (2\,M) &  0.837     &   0.9836   & 5495    &1940       &    95   &  819       & 1778       &  6192     & 4515      & 311      & 242    \\
MIParT-L (10\,M)           & 0.850     & 0.9861    & 8000      & 3003      & 112        & 1281       & 3650      & 16529      & 9852      & 440       & 336       \\
\bf MIParT-L (100\,M)      & 0.861     & 0.9878    & 10753      & 4202      & 123       & 1927       & 5450      & 31250     & 16807     & 542       & 402       \\
\bottomrule
\end{tabular} }
\end{table*}

Quark-gluon tagging is another crucial jet tagging task. Unlike the top tagging dataset, the quark-gluon dataset \cite{komiske2019energy} includes not only the kinematic features of each particle, but also particle identification information. 
This dataset allows for a more detailed categorization of particles, including specific distinctions among electrically charged and neutral hadrons, such as pions, kaons, and protons. 
Additionally, like the top tagging dataset, the quark-gluon dataset contains 2M jets, with quarks and gluons designated as the signal and background, respectively.

In Fig.~\ref{fig:6}, we showed the performance of our MIParT model compared to other popular models on the quark-gluon dataset. 
Within this dataset, the MIParT model significantly outperforms LorentzNet across all metrics, including accuracy, AUC, ${\rm Rej}_{50\%}$, and ${\rm Rej}_{30\%}$, as well as several other models. 
Moreover, only the ParT model approaches the performance of our model in several metrics, but MIParT still maintains an overall lead over ParT. 
In comparison with other models, such as PFN \cite{komiske2019energy}, ABCNet \cite{Mikuni:2020wpr}, and PCT \cite{Mikuni:2021pou}, MIParT demonstrates a substantial lead, with metrics quoted from their published results.
For the fine-tuned MIParT-L model pre-trained on the 100M JetClass dataset, a 6\% enhancement in background rejection performance is achieved, surpassing that of the fine-tuned ParT.
Detailed comparison results on the quark-gluon dataset are presented in Table~\ref{tab:3}. 
MIParT achieves the best performance across all evaluation metrics, improving background rejection power by approximately 3\% compared to ParT. 
At the same time, the background rejection of the fine-tuned MIParT-L model improved by approximately 2\% compared to the fine-tuned ParT.

Given that MIParT shares many components with ParT and differs only in the addition of the MIA blocks, the comparative results between these two models highlight the effectiveness of the MIA block. Specifically, MIParT consists of 5 MIA blocks, 5 particle attention blocks, and 2 class attention blocks, whereas ParT consists of 8 particle attention blocks and 2 class attention blocks. Thus, from the results tested on the top tagging and quark-gluon datasets, it is evident that MIParT outperforms ParT, illustrating the significant role played by the MIA block. Furthermore, the effectiveness of the particle attention blocks has already been established in the ParT paper \cite{Qu:2022mxj}, and the impact of the class attention blocks has been tested in the CaiT framework \cite{cord2021going}.

Regarding the impact of hyperparameter choices on model performance, we find that MIParT is not overly sensitive to hyperparameter settings, but is more influenced by the overall network architecture. In particular, increasing the number of MIA blocks and particle attention blocks generally leads to better performance, but at the cost of increased complexity. Architectural modifications show that placing MIA blocks before particle attention blocks is optimal. Placing MIA blocks after particle attention blocks or alternating them significantly reduces effectiveness, sometimes to the point of performing worse than ParT. We think that MIA blocks function similarly to embeddings, allowing better integration of interaction information into the jets for improved information fusion and classification.

In Table~\ref{tab:flop} we present the parameters, FLOPs (Floating Point Operations Per Second), and accuracy of various models on the top tagging and quark-gluon datasets. 
Parameters denote the number of trainable elements within a model, which indicates its capacity to learn. Conversely, more parameters generally increase the complexity of the model. 
FLOPs measure the computational complexity required to process data through the model. Reducing the number of parameters typically reduces FLOPs, simplifying the model and making it more computationally efficient.

However, reducing the number of parameters to reduce FLOPs usually results in lower accuracy. In contrast, our MIParT model has only 30\% of the parameters and 53\% of the FLOPs of the ParT model, significantly reducing model complexity. 
Despite this reduction, there is no compromise in accuracy; in fact, accuracy improves on both top tagging and quark-gluon datasets.
For the fine-tuned version of MIParT-L, the parameters and FLOPs are comparable to those of the ParT model, but with a slight improvement in accuracy.

In Table~\ref{tab:jc}, we present comparative performance of various models on different sizes of the JetClass dataset. we displaced the results for the MIParT-L model alongside ParticleNet \cite{Qu:2019gqs} and ParT \cite{Qu:2022mxj} across 2M, 10M, and 100M JetClass datasets. 
We observe that as the dataset size increases, the performance of the models improves. Specifically, MIParT-L and ParT exhibit nearly identical effectiveness on very large datasets, surpassing that of ParticleNet.
In addition, our evaluation of models on the JetClass dataset serves to test the ability of MIParT to generalize across different classification tasks. The JetClass datasets represent a more complex classification challenge, aimed at identifying Higgs boson decays to charm quarks. Our MIParT model shows remarkable stability on this task, highlighting its generalization capabilities.

Here, we discuss the improvements attributed to the pre-training performed on the JetClass dataset, with subsequent performance improvements observed on the top tagging and quark-gluon datasets. These three jet tagging tasks differ in their objectives: the JetClass dataset focuses on identifying Lorentz boosted $W$, $Z$, Higgs bosons and top quarks, the top tagging dataset aims to identify top quarks, and the quark-gluon dataset aims to distinguish between quark and gluon jets. The improvements across such diverse tasks suggest that MIParT has learned more generalized jet properties during the pre-training phase. These characteristics are effectively transferable to other tasks, demonstrating the model's robustness and adaptability to different jet identification challenges. This capability highlights the potential of pre-trained models to improve performance in a wide range of applications by capturing and exploiting general features applicable to multiple scenarios.

Regarding the interpretability of MIParT, it is important to acknowledge that as a model based on the transformer neural network architecture, its interpretability remains limited, similar to many neural networks currently in use. Despite these interpretability challenges, the CMS collaboration has successfully used the graph neural network ParticleNet \cite{Qu:2019gqs}, another model that lacks full interpretability, to search for Higgs boson decay to charm quarks \cite{CMS:2022psv}. This success underscores that the lack of interpretability does not prevent the use of neural network models in particle physics experiments. In fact, ParticleNet, which functions as a non-interpretable ``black box" model, has already begun to play a significant role in particle experiments, demonstrating that the non-interpretable nature of these models should not be a barrier to their use in advancing scientific discovery.

\section{Conclusion}
\label{sec:con}

In this paper, we propose a novel deep learning approach for jet tagging, MIParT. MIParT increases the dimensionality of particle interaction embeddings through More-Interaction Attention (MIA) to better utilize particle interaction inputs. We tested our model on two popular datasets and compared it with other models:
\begin{itemize}
    \item \textbf{On the Top Tagging Dataset:} The MIParT model achieved accuracy and AUC metrics nearly identical to those of LorentzNet, and its ${\rm Rej}_{50\%}$ and ${\rm Rej}_{30\%}$ metrics are comparable within the error range to LorentzNet. And a series of Lorentz-equivariant methods demonstrated performance similar to that of LorentzNet. The MIParT model significantly outperformed ParT in the top tagging benchmark, achieving approximately 25\% better background rejection at a 30\% signal efficiency. Among the models evaluated, MIParT, along with LorentzNet and several other Lorentz-equivariant based models, ranks at the top tier, showcasing some of the most robust performances. For the fine-tuned MIParT-L model that is pre-trained on the 100M JetClass dataset, a 39\% enhancement in background rejection performance was achieved, comparable to that of the fine-tuned ParT. 

    \item \textbf{On the Quark-gluon Dataset:} The MIParT model significantly outperforms LorentzNet across all metrics, including accuracy, AUC, ${\rm Rej}_{50\%}$, and ${\rm Rej}_{30\%}$, as well as several other models. MIParT achieved the best performance across all evaluation metrics, improving background rejection power by approximately 3\% compared to ParT. 
    For the fine-tuned MIParT-L model, background rejection performance improved by 6\%, surpassing that of the fine-tuned ParT. Specifically, the background rejection of fine-tuned MIParT-L improved by an additional 2\% compared to the fine-tuned ParT.
    
\end{itemize}

Overall, MIParT outperformed ParT on both the top tagging and quark-gluon tagging tasks while also exhibiting lower computational complexity and fewer parameters. Previously, it was generally assumed that transformer-based models required large-scale dataset pre-training to achieve optimal results. Our MIParT model demonstrates that with higher-dimensional particle interaction embeddings, top-tier performance can be achieved without pre-training on large datasets, even surpassing ParT.

Furthermore, as pre-training ParT on the larger multi-class JetClass dataset and subsequently fine-tuning it on the top tagging dataset can enhance performance, we have applied this approach to MIParT-L in this work.
We find that MIParT-L can further capitalize on the knowledge from large datasets, showing superior capabilities after fine-tuning. Specifically, it performs better on the quark-gluon dataset than the fine-tuned ParT.
Finding more efficient ways to fine-tune a base Transformer model will be especially helpful for future experiments when generic and foundation models are deployed, and downstream application tasks are varied.
Moreover, MIParT is not limited to jet tagging but can also be applied to event identification, which could be immensely helpful in the search for new physics signals.

\begin{acknowledgments}
The work of K. Wang and J. Zhu is supported by the National Natural Science Foundation of China (NNSFC) under grant Nos. 12275066 and 11605123.
\end{acknowledgments}

\appendix





\bibliographystyle{apsrev4-1}
\bibliography{apssamp}

\begin{thebibliography}{49}%
\makeatletter
\providecommand \@ifxundefined [1]{%
 \@ifx{#1\undefined}
}%
\providecommand \@ifnum [1]{%
 \ifnum #1\expandafter \@firstoftwo
 \else \expandafter \@secondoftwo
 \fi
}%
\providecommand \@ifx [1]{%
 \ifx #1\expandafter \@firstoftwo
 \else \expandafter \@secondoftwo
 \fi
}%
\providecommand \natexlab [1]{#1}%
\providecommand \enquote  [1]{``#1''}%
\providecommand \bibnamefont  [1]{#1}%
\providecommand \bibfnamefont [1]{#1}%
\providecommand \citenamefont [1]{#1}%
\providecommand \href@noop [0]{\@secondoftwo}%
\providecommand \href [0]{\begingroup \@sanitize@url \@href}%
\providecommand \@href[1]{\@@startlink{#1}\@@href}%
\providecommand \@@href[1]{\endgroup#1\@@endlink}%
\providecommand \@sanitize@url [0]{\catcode `\\12\catcode `\$12\catcode `\&12\catcode `\#12\catcode `\^12\catcode `\_12\catcode `\%12\relax}%
\providecommand \@@startlink[1]{}%
\providecommand \@@endlink[0]{}%
\providecommand \url  [0]{\begingroup\@sanitize@url \@url }%
\providecommand \@url [1]{\endgroup\@href {#1}{\urlprefix }}%
\providecommand \urlprefix  [0]{URL }%
\providecommand \Eprint [0]{\href }%
\providecommand \doibase [0]{http://dx.doi.org/}%
\providecommand \selectlanguage [0]{\@gobble}%
\providecommand \bibinfo  [0]{\@secondoftwo}%
\providecommand \bibfield  [0]{\@secondoftwo}%
\providecommand \translation [1]{[#1]}%
\providecommand \BibitemOpen [0]{}%
\providecommand \bibitemStop [0]{}%
\providecommand \bibitemNoStop [0]{.\EOS\space}%
\providecommand \EOS [0]{\spacefactor3000\relax}%
\providecommand \BibitemShut  [1]{\csname bibitem#1\endcsname}%
\let\auto@bib@innerbib\@empty
\bibitem [{\citenamefont {Larkoski}\ \emph {et~al.}(2020)\citenamefont {Larkoski}, \citenamefont {Moult},\ and\ \citenamefont {Nachman}}]{Larkoski:2017jix}%
  \BibitemOpen
  \bibfield  {author} {\bibinfo {author} {\bibfnamefont {A.~J.}\ \bibnamefont {Larkoski}}, \bibinfo {author} {\bibfnamefont {I.}~\bibnamefont {Moult}}, \ and\ \bibinfo {author} {\bibfnamefont {B.}~\bibnamefont {Nachman}},\ }\href {\doibase 10.1016/j.physrep.2019.11.001} {\bibfield  {journal} {\bibinfo  {journal} {Phys. Rept.}\ }\textbf {\bibinfo {volume} {841}},\ \bibinfo {pages} {1} (\bibinfo {year} {2020})},\ \Eprint {http://arxiv.org/abs/1709.04464} {arXiv:1709.04464 [hep-ph]} \BibitemShut {NoStop}%
\bibitem [{\citenamefont {Feickert}\ and\ \citenamefont {Nachman}(2021)}]{Feickert:2021ajf}%
  \BibitemOpen
  \bibfield  {author} {\bibinfo {author} {\bibfnamefont {M.}~\bibnamefont {Feickert}}\ and\ \bibinfo {author} {\bibfnamefont {B.}~\bibnamefont {Nachman}},\ }\href@noop {} {\  (\bibinfo {year} {2021})},\ \Eprint {http://arxiv.org/abs/2102.02770} {arXiv:2102.02770 [hep-ph]} \BibitemShut {NoStop}%
\bibitem [{\citenamefont {Gallicchio}\ and\ \citenamefont {Schwartz}(2011)}]{Gallicchio:2011xq}%
  \BibitemOpen
  \bibfield  {author} {\bibinfo {author} {\bibfnamefont {J.}~\bibnamefont {Gallicchio}}\ and\ \bibinfo {author} {\bibfnamefont {M.~D.}\ \bibnamefont {Schwartz}},\ }\href {\doibase 10.1103/PhysRevLett.107.172001} {\bibfield  {journal} {\bibinfo  {journal} {Phys. Rev. Lett.}\ }\textbf {\bibinfo {volume} {107}},\ \bibinfo {pages} {172001} (\bibinfo {year} {2011})},\ \Eprint {http://arxiv.org/abs/1106.3076} {arXiv:1106.3076 [hep-ph]} \BibitemShut {NoStop}%
\bibitem [{\citenamefont {Gallicchio}\ and\ \citenamefont {Schwartz}(2013)}]{Gallicchio:2012ez}%
  \BibitemOpen
  \bibfield  {author} {\bibinfo {author} {\bibfnamefont {J.}~\bibnamefont {Gallicchio}}\ and\ \bibinfo {author} {\bibfnamefont {M.~D.}\ \bibnamefont {Schwartz}},\ }\href {\doibase 10.1007/JHEP04(2013)090} {\bibfield  {journal} {\bibinfo  {journal} {JHEP}\ }\textbf {\bibinfo {volume} {04}},\ \bibinfo {pages} {090} (\bibinfo {year} {2013})},\ \Eprint {http://arxiv.org/abs/1211.7038} {arXiv:1211.7038 [hep-ph]} \BibitemShut {NoStop}%
\bibitem [{\citenamefont {Larkoski}\ \emph {et~al.}(2014)\citenamefont {Larkoski}, \citenamefont {Thaler},\ and\ \citenamefont {Waalewijn}}]{Larkoski:2014pca}%
  \BibitemOpen
  \bibfield  {author} {\bibinfo {author} {\bibfnamefont {A.~J.}\ \bibnamefont {Larkoski}}, \bibinfo {author} {\bibfnamefont {J.}~\bibnamefont {Thaler}}, \ and\ \bibinfo {author} {\bibfnamefont {W.~J.}\ \bibnamefont {Waalewijn}},\ }\href {\doibase 10.1007/JHEP11(2014)129} {\bibfield  {journal} {\bibinfo  {journal} {JHEP}\ }\textbf {\bibinfo {volume} {11}},\ \bibinfo {pages} {129} (\bibinfo {year} {2014})},\ \Eprint {http://arxiv.org/abs/1408.3122} {arXiv:1408.3122 [hep-ph]} \BibitemShut {NoStop}%
\bibitem [{\citenamefont {Bhattacherjee}\ \emph {et~al.}(2015)\citenamefont {Bhattacherjee}, \citenamefont {Mukhopadhyay}, \citenamefont {Nojiri}, \citenamefont {Sakaki},\ and\ \citenamefont {Webber}}]{Bhattacherjee:2015psa}%
  \BibitemOpen
  \bibfield  {author} {\bibinfo {author} {\bibfnamefont {B.}~\bibnamefont {Bhattacherjee}}, \bibinfo {author} {\bibfnamefont {S.}~\bibnamefont {Mukhopadhyay}}, \bibinfo {author} {\bibfnamefont {M.~M.}\ \bibnamefont {Nojiri}}, \bibinfo {author} {\bibfnamefont {Y.}~\bibnamefont {Sakaki}}, \ and\ \bibinfo {author} {\bibfnamefont {B.~R.}\ \bibnamefont {Webber}},\ }\href {\doibase 10.1007/JHEP04(2015)131} {\bibfield  {journal} {\bibinfo  {journal} {JHEP}\ }\textbf {\bibinfo {volume} {04}},\ \bibinfo {pages} {131} (\bibinfo {year} {2015})},\ \Eprint {http://arxiv.org/abs/1501.04794} {arXiv:1501.04794 [hep-ph]} \BibitemShut {NoStop}%
\bibitem [{\citenamefont {Ferreira~de Lima}\ \emph {et~al.}(2017)\citenamefont {Ferreira~de Lima}, \citenamefont {Petrov}, \citenamefont {Soper},\ and\ \citenamefont {Spannowsky}}]{FerreiradeLima:2016gcz}%
  \BibitemOpen
  \bibfield  {author} {\bibinfo {author} {\bibfnamefont {D.}~\bibnamefont {Ferreira~de Lima}}, \bibinfo {author} {\bibfnamefont {P.}~\bibnamefont {Petrov}}, \bibinfo {author} {\bibfnamefont {D.}~\bibnamefont {Soper}}, \ and\ \bibinfo {author} {\bibfnamefont {M.}~\bibnamefont {Spannowsky}},\ }\href {\doibase 10.1103/PhysRevD.95.034001} {\bibfield  {journal} {\bibinfo  {journal} {Phys. Rev. D}\ }\textbf {\bibinfo {volume} {95}},\ \bibinfo {pages} {034001} (\bibinfo {year} {2017})},\ \Eprint {http://arxiv.org/abs/1607.06031} {arXiv:1607.06031 [hep-ph]} \BibitemShut {NoStop}%
\bibitem [{\citenamefont {Gras}\ \emph {et~al.}(2017)\citenamefont {Gras}, \citenamefont {H\"oche}, \citenamefont {Kar}, \citenamefont {Larkoski}, \citenamefont {L\"onnblad}, \citenamefont {Pl\"atzer}, \citenamefont {Si\'odmok}, \citenamefont {Skands}, \citenamefont {Soyez},\ and\ \citenamefont {Thaler}}]{Gras:2017jty}%
  \BibitemOpen
  \bibfield  {author} {\bibinfo {author} {\bibfnamefont {P.}~\bibnamefont {Gras}}, \bibinfo {author} {\bibfnamefont {S.}~\bibnamefont {H\"oche}}, \bibinfo {author} {\bibfnamefont {D.}~\bibnamefont {Kar}}, \bibinfo {author} {\bibfnamefont {A.}~\bibnamefont {Larkoski}}, \bibinfo {author} {\bibfnamefont {L.}~\bibnamefont {L\"onnblad}}, \bibinfo {author} {\bibfnamefont {S.}~\bibnamefont {Pl\"atzer}}, \bibinfo {author} {\bibfnamefont {A.}~\bibnamefont {Si\'odmok}}, \bibinfo {author} {\bibfnamefont {P.}~\bibnamefont {Skands}}, \bibinfo {author} {\bibfnamefont {G.}~\bibnamefont {Soyez}}, \ and\ \bibinfo {author} {\bibfnamefont {J.}~\bibnamefont {Thaler}},\ }\href {\doibase 10.1007/JHEP07(2017)091} {\bibfield  {journal} {\bibinfo  {journal} {JHEP}\ }\textbf {\bibinfo {volume} {07}},\ \bibinfo {pages} {091} (\bibinfo {year} {2017})},\ \Eprint {http://arxiv.org/abs/1704.03878} {arXiv:1704.03878 [hep-ph]} \BibitemShut {NoStop}%
\bibitem [{\citenamefont {Frye}\ \emph {et~al.}(2017)\citenamefont {Frye}, \citenamefont {Larkoski}, \citenamefont {Thaler},\ and\ \citenamefont {Zhou}}]{Frye:2017yrw}%
  \BibitemOpen
  \bibfield  {author} {\bibinfo {author} {\bibfnamefont {C.}~\bibnamefont {Frye}}, \bibinfo {author} {\bibfnamefont {A.~J.}\ \bibnamefont {Larkoski}}, \bibinfo {author} {\bibfnamefont {J.}~\bibnamefont {Thaler}}, \ and\ \bibinfo {author} {\bibfnamefont {K.}~\bibnamefont {Zhou}},\ }\href {\doibase 10.1007/JHEP09(2017)083} {\bibfield  {journal} {\bibinfo  {journal} {JHEP}\ }\textbf {\bibinfo {volume} {09}},\ \bibinfo {pages} {083} (\bibinfo {year} {2017})},\ \Eprint {http://arxiv.org/abs/1704.06266} {arXiv:1704.06266 [hep-ph]} \BibitemShut {NoStop}%
\bibitem [{\citenamefont {Cogan}\ \emph {et~al.}(2015)\citenamefont {Cogan}, \citenamefont {Kagan}, \citenamefont {Strauss},\ and\ \citenamefont {Schwarztman}}]{Cogan:2014oua}%
  \BibitemOpen
  \bibfield  {author} {\bibinfo {author} {\bibfnamefont {J.}~\bibnamefont {Cogan}}, \bibinfo {author} {\bibfnamefont {M.}~\bibnamefont {Kagan}}, \bibinfo {author} {\bibfnamefont {E.}~\bibnamefont {Strauss}}, \ and\ \bibinfo {author} {\bibfnamefont {A.}~\bibnamefont {Schwarztman}},\ }\href {\doibase 10.1007/JHEP02(2015)118} {\bibfield  {journal} {\bibinfo  {journal} {JHEP}\ }\textbf {\bibinfo {volume} {02}},\ \bibinfo {pages} {118} (\bibinfo {year} {2015})},\ \Eprint {http://arxiv.org/abs/1407.5675} {arXiv:1407.5675 [hep-ph]} \BibitemShut {NoStop}%
\bibitem [{\citenamefont {Almeida}\ \emph {et~al.}(2015)\citenamefont {Almeida}, \citenamefont {Backovi\'c}, \citenamefont {Cliche}, \citenamefont {Lee},\ and\ \citenamefont {Perelstein}}]{Almeida:2015jua}%
  \BibitemOpen
  \bibfield  {author} {\bibinfo {author} {\bibfnamefont {L.~G.}\ \bibnamefont {Almeida}}, \bibinfo {author} {\bibfnamefont {M.}~\bibnamefont {Backovi\'c}}, \bibinfo {author} {\bibfnamefont {M.}~\bibnamefont {Cliche}}, \bibinfo {author} {\bibfnamefont {S.~J.}\ \bibnamefont {Lee}}, \ and\ \bibinfo {author} {\bibfnamefont {M.}~\bibnamefont {Perelstein}},\ }\href {\doibase 10.1007/JHEP07(2015)086} {\bibfield  {journal} {\bibinfo  {journal} {JHEP}\ }\textbf {\bibinfo {volume} {07}},\ \bibinfo {pages} {086} (\bibinfo {year} {2015})},\ \Eprint {http://arxiv.org/abs/1501.05968} {arXiv:1501.05968 [hep-ph]} \BibitemShut {NoStop}%
\bibitem [{\citenamefont {de~Oliveira}\ \emph {et~al.}(2016)\citenamefont {de~Oliveira}, \citenamefont {Kagan}, \citenamefont {Mackey}, \citenamefont {Nachman},\ and\ \citenamefont {Schwartzman}}]{deOliveira:2015xxd}%
  \BibitemOpen
  \bibfield  {author} {\bibinfo {author} {\bibfnamefont {L.}~\bibnamefont {de~Oliveira}}, \bibinfo {author} {\bibfnamefont {M.}~\bibnamefont {Kagan}}, \bibinfo {author} {\bibfnamefont {L.}~\bibnamefont {Mackey}}, \bibinfo {author} {\bibfnamefont {B.}~\bibnamefont {Nachman}}, \ and\ \bibinfo {author} {\bibfnamefont {A.}~\bibnamefont {Schwartzman}},\ }\href {\doibase 10.1007/JHEP07(2016)069} {\bibfield  {journal} {\bibinfo  {journal} {JHEP}\ }\textbf {\bibinfo {volume} {07}},\ \bibinfo {pages} {069} (\bibinfo {year} {2016})},\ \Eprint {http://arxiv.org/abs/1511.05190} {arXiv:1511.05190 [hep-ph]} \BibitemShut {NoStop}%
\bibitem [{\citenamefont {Komiske}\ \emph {et~al.}(2017)\citenamefont {Komiske}, \citenamefont {Metodiev},\ and\ \citenamefont {Schwartz}}]{Komiske:2016rsd}%
  \BibitemOpen
  \bibfield  {author} {\bibinfo {author} {\bibfnamefont {P.~T.}\ \bibnamefont {Komiske}}, \bibinfo {author} {\bibfnamefont {E.~M.}\ \bibnamefont {Metodiev}}, \ and\ \bibinfo {author} {\bibfnamefont {M.~D.}\ \bibnamefont {Schwartz}},\ }\href {\doibase 10.1007/JHEP01(2017)110} {\bibfield  {journal} {\bibinfo  {journal} {JHEP}\ }\textbf {\bibinfo {volume} {01}},\ \bibinfo {pages} {110} (\bibinfo {year} {2017})},\ \Eprint {http://arxiv.org/abs/1612.01551} {arXiv:1612.01551 [hep-ph]} \BibitemShut {NoStop}%
\bibitem [{\citenamefont {Kasieczka}\ \emph {et~al.}(2017)\citenamefont {Kasieczka}, \citenamefont {Plehn}, \citenamefont {Russell},\ and\ \citenamefont {Schell}}]{Kasieczka:2017nvn}%
  \BibitemOpen
  \bibfield  {author} {\bibinfo {author} {\bibfnamefont {G.}~\bibnamefont {Kasieczka}}, \bibinfo {author} {\bibfnamefont {T.}~\bibnamefont {Plehn}}, \bibinfo {author} {\bibfnamefont {M.}~\bibnamefont {Russell}}, \ and\ \bibinfo {author} {\bibfnamefont {T.}~\bibnamefont {Schell}},\ }\href {\doibase 10.1007/JHEP05(2017)006} {\bibfield  {journal} {\bibinfo  {journal} {JHEP}\ }\textbf {\bibinfo {volume} {05}},\ \bibinfo {pages} {006} (\bibinfo {year} {2017})},\ \Eprint {http://arxiv.org/abs/1701.08784} {arXiv:1701.08784 [hep-ph]} \BibitemShut {NoStop}%
\bibitem [{\citenamefont {Macaluso}\ and\ \citenamefont {Shih}(2018)}]{Macaluso:2018tck}%
  \BibitemOpen
  \bibfield  {author} {\bibinfo {author} {\bibfnamefont {S.}~\bibnamefont {Macaluso}}\ and\ \bibinfo {author} {\bibfnamefont {D.}~\bibnamefont {Shih}},\ }\href {\doibase 10.1007/JHEP10(2018)121} {\bibfield  {journal} {\bibinfo  {journal} {JHEP}\ }\textbf {\bibinfo {volume} {10}},\ \bibinfo {pages} {121} (\bibinfo {year} {2018})},\ \Eprint {http://arxiv.org/abs/1803.00107} {arXiv:1803.00107 [hep-ph]} \BibitemShut {NoStop}%
\bibitem [{\citenamefont {Butter}\ \emph {et~al.}(2019)\citenamefont {Butter} \emph {et~al.}}]{Kasieczka:2019dbj}%
  \BibitemOpen
  \bibfield  {author} {\bibinfo {author} {\bibfnamefont {A.}~\bibnamefont {Butter}} \emph {et~al.},\ }\href {\doibase 10.21468/SciPostPhys.7.1.014} {\bibfield  {journal} {\bibinfo  {journal} {SciPost Phys.}\ }\textbf {\bibinfo {volume} {7}},\ \bibinfo {pages} {014} (\bibinfo {year} {2019})},\ \Eprint {http://arxiv.org/abs/1902.09914} {arXiv:1902.09914 [hep-ph]} \BibitemShut {NoStop}%
\bibitem [{\citenamefont {Kagan}(2020)}]{Kagan:2020yrm}%
  \BibitemOpen
  \bibfield  {author} {\bibinfo {author} {\bibfnamefont {M.}~\bibnamefont {Kagan}},\ }\href@noop {} {\  (\bibinfo {year} {2020})},\ \Eprint {http://arxiv.org/abs/2012.09719} {arXiv:2012.09719 [physics.data-an]} \BibitemShut {NoStop}%
\bibitem [{\citenamefont {de~Lima}(2021)}]{deLima:2021fwm}%
  \BibitemOpen
  \bibfield  {author} {\bibinfo {author} {\bibfnamefont {R.~T.}\ \bibnamefont {de~Lima}},\ }\href@noop {} {\  (\bibinfo {year} {2021})},\ \Eprint {http://arxiv.org/abs/2102.06128} {arXiv:2102.06128 [physics.data-an]} \BibitemShut {NoStop}%
\bibitem [{\citenamefont {Kheddar}\ \emph {et~al.}(2024)\citenamefont {Kheddar}, \citenamefont {Himeur}, \citenamefont {Amira},\ and\ \citenamefont {Soualah}}]{Kheddar:2024osf}%
  \BibitemOpen
  \bibfield  {author} {\bibinfo {author} {\bibfnamefont {H.}~\bibnamefont {Kheddar}}, \bibinfo {author} {\bibfnamefont {Y.}~\bibnamefont {Himeur}}, \bibinfo {author} {\bibfnamefont {A.}~\bibnamefont {Amira}}, \ and\ \bibinfo {author} {\bibfnamefont {R.}~\bibnamefont {Soualah}},\ }\href@noop {} {\  (\bibinfo {year} {2024})},\ \Eprint {http://arxiv.org/abs/2403.11934} {arXiv:2403.11934 [hep-ph]} \BibitemShut {NoStop}%
\bibitem [{\citenamefont {Mondal}\ and\ \citenamefont {Mastrolorenzo}(2024)}]{Mondal:2024nsa}%
  \BibitemOpen
  \bibfield  {author} {\bibinfo {author} {\bibfnamefont {S.}~\bibnamefont {Mondal}}\ and\ \bibinfo {author} {\bibfnamefont {L.}~\bibnamefont {Mastrolorenzo}},\ }\href@noop {} {\  (\bibinfo {year} {2024})},\ \Eprint {http://arxiv.org/abs/2404.01071} {arXiv:2404.01071 [hep-ex]} \BibitemShut {NoStop}%
\bibitem [{\citenamefont {Pumplin}(1991)}]{Pumplin:1991kc}%
  \BibitemOpen
  \bibfield  {author} {\bibinfo {author} {\bibfnamefont {J.}~\bibnamefont {Pumplin}},\ }\href {\doibase 10.1103/PhysRevD.44.2025} {\bibfield  {journal} {\bibinfo  {journal} {Phys. Rev. D}\ }\textbf {\bibinfo {volume} {44}},\ \bibinfo {pages} {2025} (\bibinfo {year} {1991})}\BibitemShut {NoStop}%
\bibitem [{\citenamefont {Lin}\ \emph {et~al.}(2018)\citenamefont {Lin}, \citenamefont {Freytsis}, \citenamefont {Moult},\ and\ \citenamefont {Nachman}}]{Lin:2018cin}%
  \BibitemOpen
  \bibfield  {author} {\bibinfo {author} {\bibfnamefont {J.}~\bibnamefont {Lin}}, \bibinfo {author} {\bibfnamefont {M.}~\bibnamefont {Freytsis}}, \bibinfo {author} {\bibfnamefont {I.}~\bibnamefont {Moult}}, \ and\ \bibinfo {author} {\bibfnamefont {B.}~\bibnamefont {Nachman}},\ }\href {\doibase 10.1007/JHEP10(2018)101} {\bibfield  {journal} {\bibinfo  {journal} {JHEP}\ }\textbf {\bibinfo {volume} {10}},\ \bibinfo {pages} {101} (\bibinfo {year} {2018})},\ \Eprint {http://arxiv.org/abs/1807.10768} {arXiv:1807.10768 [hep-ph]} \BibitemShut {NoStop}%
\bibitem [{\citenamefont {Guest}\ \emph {et~al.}(2016)\citenamefont {Guest}, \citenamefont {Collado}, \citenamefont {Baldi}, \citenamefont {Hsu}, \citenamefont {Urban},\ and\ \citenamefont {Whiteson}}]{Guest:2016iqz}%
  \BibitemOpen
  \bibfield  {author} {\bibinfo {author} {\bibfnamefont {D.}~\bibnamefont {Guest}}, \bibinfo {author} {\bibfnamefont {J.}~\bibnamefont {Collado}}, \bibinfo {author} {\bibfnamefont {P.}~\bibnamefont {Baldi}}, \bibinfo {author} {\bibfnamefont {S.-C.}\ \bibnamefont {Hsu}}, \bibinfo {author} {\bibfnamefont {G.}~\bibnamefont {Urban}}, \ and\ \bibinfo {author} {\bibfnamefont {D.}~\bibnamefont {Whiteson}},\ }\href {\doibase 10.1103/PhysRevD.94.112002} {\bibfield  {journal} {\bibinfo  {journal} {Phys. Rev. D}\ }\textbf {\bibinfo {volume} {94}},\ \bibinfo {pages} {112002} (\bibinfo {year} {2016})},\ \Eprint {http://arxiv.org/abs/1607.08633} {arXiv:1607.08633 [hep-ex]} \BibitemShut {NoStop}%
\bibitem [{\citenamefont {Bols}\ \emph {et~al.}(2020)\citenamefont {Bols}, \citenamefont {Kieseler}, \citenamefont {Verzetti}, \citenamefont {Stoye},\ and\ \citenamefont {Stakia}}]{Bols:2020bkb}%
  \BibitemOpen
  \bibfield  {author} {\bibinfo {author} {\bibfnamefont {E.}~\bibnamefont {Bols}}, \bibinfo {author} {\bibfnamefont {J.}~\bibnamefont {Kieseler}}, \bibinfo {author} {\bibfnamefont {M.}~\bibnamefont {Verzetti}}, \bibinfo {author} {\bibfnamefont {M.}~\bibnamefont {Stoye}}, \ and\ \bibinfo {author} {\bibfnamefont {A.}~\bibnamefont {Stakia}},\ }\href {\doibase 10.1088/1748-0221/15/12/P12012} {\bibfield  {journal} {\bibinfo  {journal} {JINST}\ }\textbf {\bibinfo {volume} {15}},\ \bibinfo {pages} {P12012} (\bibinfo {year} {2020})},\ \Eprint {http://arxiv.org/abs/2008.10519} {arXiv:2008.10519 [hep-ex]} \BibitemShut {NoStop}%
\bibitem [{\citenamefont {Louppe}\ \emph {et~al.}(2019)\citenamefont {Louppe}, \citenamefont {Cho}, \citenamefont {Becot},\ and\ \citenamefont {Cranmer}}]{Louppe:2017ipp}%
  \BibitemOpen
  \bibfield  {author} {\bibinfo {author} {\bibfnamefont {G.}~\bibnamefont {Louppe}}, \bibinfo {author} {\bibfnamefont {K.}~\bibnamefont {Cho}}, \bibinfo {author} {\bibfnamefont {C.}~\bibnamefont {Becot}}, \ and\ \bibinfo {author} {\bibfnamefont {K.}~\bibnamefont {Cranmer}},\ }\href {\doibase 10.1007/JHEP01(2019)057} {\bibfield  {journal} {\bibinfo  {journal} {JHEP}\ }\textbf {\bibinfo {volume} {01}},\ \bibinfo {pages} {057} (\bibinfo {year} {2019})},\ \Eprint {http://arxiv.org/abs/1702.00748} {arXiv:1702.00748 [hep-ph]} \BibitemShut {NoStop}%
\bibitem [{\citenamefont {Cheng}(2018)}]{Cheng:2017rdo}%
  \BibitemOpen
  \bibfield  {author} {\bibinfo {author} {\bibfnamefont {T.}~\bibnamefont {Cheng}},\ }\href {\doibase 10.1007/s41781-018-0007-y} {\bibfield  {journal} {\bibinfo  {journal} {Comput. Softw. Big Sci.}\ }\textbf {\bibinfo {volume} {2}},\ \bibinfo {pages} {3} (\bibinfo {year} {2018})},\ \Eprint {http://arxiv.org/abs/1711.02633} {arXiv:1711.02633 [hep-ph]} \BibitemShut {NoStop}%
\bibitem [{\citenamefont {Henrion}\ \emph {et~al.}(2017)\citenamefont {Henrion}, \citenamefont {Cranmer}, \citenamefont {Bruna}, \citenamefont {Cho}, \citenamefont {Brehmer}, \citenamefont {Louppe},\ and\ \citenamefont {Rochette}}]{Henrion:DLPS2017}%
  \BibitemOpen
  \bibfield  {author} {\bibinfo {author} {\bibfnamefont {I.}~\bibnamefont {Henrion}}, \bibinfo {author} {\bibfnamefont {K.}~\bibnamefont {Cranmer}}, \bibinfo {author} {\bibfnamefont {J.}~\bibnamefont {Bruna}}, \bibinfo {author} {\bibfnamefont {K.}~\bibnamefont {Cho}}, \bibinfo {author} {\bibfnamefont {J.}~\bibnamefont {Brehmer}}, \bibinfo {author} {\bibfnamefont {G.}~\bibnamefont {Louppe}}, \ and\ \bibinfo {author} {\bibfnamefont {G.}~\bibnamefont {Rochette}},\ }\href {https://dl4physicalsciences.github.io/files/nips_dlps_2017_29.pdf} {\bibfield  {journal} {\bibinfo  {journal} {{Proceedings of the Deep Learning for Physical Sciences Workshop at NIPS (2017)}}\ } (\bibinfo {year} {2017})}\BibitemShut {NoStop}%
\bibitem [{\citenamefont {Abdughani}\ \emph {et~al.}(2019)\citenamefont {Abdughani}, \citenamefont {Ren}, \citenamefont {Wu},\ and\ \citenamefont {Yang}}]{Abdughani:2018wrw}%
  \BibitemOpen
  \bibfield  {author} {\bibinfo {author} {\bibfnamefont {M.}~\bibnamefont {Abdughani}}, \bibinfo {author} {\bibfnamefont {J.}~\bibnamefont {Ren}}, \bibinfo {author} {\bibfnamefont {L.}~\bibnamefont {Wu}}, \ and\ \bibinfo {author} {\bibfnamefont {J.~M.}\ \bibnamefont {Yang}},\ }\href {\doibase 10.1007/JHEP08(2019)055} {\bibfield  {journal} {\bibinfo  {journal} {JHEP}\ }\textbf {\bibinfo {volume} {08}},\ \bibinfo {pages} {055} (\bibinfo {year} {2019})},\ \Eprint {http://arxiv.org/abs/1807.09088} {arXiv:1807.09088 [hep-ph]} \BibitemShut {NoStop}%
\bibitem [{\citenamefont {Arjona~Martínez}\ \emph {et~al.}(2019)\citenamefont {Arjona~Martínez}, \citenamefont {Cerri}, \citenamefont {Pierini}, \citenamefont {Spiropulu},\ and\ \citenamefont {Vlimant}}]{Martinez:2018fwc}%
  \BibitemOpen
  \bibfield  {author} {\bibinfo {author} {\bibfnamefont {J.}~\bibnamefont {Arjona~Martínez}}, \bibinfo {author} {\bibfnamefont {O.}~\bibnamefont {Cerri}}, \bibinfo {author} {\bibfnamefont {M.}~\bibnamefont {Pierini}}, \bibinfo {author} {\bibfnamefont {M.}~\bibnamefont {Spiropulu}}, \ and\ \bibinfo {author} {\bibfnamefont {J.-R.}\ \bibnamefont {Vlimant}},\ }\href {\doibase 10.1140/epjp/i2019-12710-3} {\bibfield  {journal} {\bibinfo  {journal} {Eur. Phys. J. Plus}\ }\textbf {\bibinfo {volume} {134}},\ \bibinfo {pages} {333} (\bibinfo {year} {2019})},\ \Eprint {http://arxiv.org/abs/1810.07988} {arXiv:1810.07988 [hep-ph]} \BibitemShut {NoStop}%
\bibitem [{\citenamefont {Ren}\ \emph {et~al.}(2020)\citenamefont {Ren}, \citenamefont {Wu},\ and\ \citenamefont {Yang}}]{Ren:2019xhp}%
  \BibitemOpen
  \bibfield  {author} {\bibinfo {author} {\bibfnamefont {J.}~\bibnamefont {Ren}}, \bibinfo {author} {\bibfnamefont {L.}~\bibnamefont {Wu}}, \ and\ \bibinfo {author} {\bibfnamefont {J.~M.}\ \bibnamefont {Yang}},\ }\href {\doibase 10.1016/j.physletb.2020.135198} {\bibfield  {journal} {\bibinfo  {journal} {Phys. Lett. B}\ }\textbf {\bibinfo {volume} {802}},\ \bibinfo {pages} {135198} (\bibinfo {year} {2020})},\ \Eprint {http://arxiv.org/abs/1901.05627} {arXiv:1901.05627 [hep-ph]} \BibitemShut {NoStop}%
\bibitem [{\citenamefont {Ju}\ \emph {et~al.}(2020)\citenamefont {Ju} \emph {et~al.}}]{Ju:2020xty}%
  \BibitemOpen
  \bibfield  {author} {\bibinfo {author} {\bibfnamefont {X.}~\bibnamefont {Ju}} \emph {et~al.},\ }\href@noop {} {\bibfield  {journal} {\bibinfo  {journal} {{33rd Annual Conference on Neural Information Processing Systems}}\ } (\bibinfo {year} {2020})},\ \Eprint {http://arxiv.org/abs/2003.11603} {arXiv:2003.11603 [physics.ins-det]} \BibitemShut {NoStop}%
\bibitem [{\citenamefont {Komiske}\ \emph {et~al.}(2019{\natexlab{a}})\citenamefont {Komiske}, \citenamefont {Metodiev},\ and\ \citenamefont {Thaler}}]{Komiske:2018cqr}%
  \BibitemOpen
  \bibfield  {author} {\bibinfo {author} {\bibfnamefont {P.~T.}\ \bibnamefont {Komiske}}, \bibinfo {author} {\bibfnamefont {E.~M.}\ \bibnamefont {Metodiev}}, \ and\ \bibinfo {author} {\bibfnamefont {J.}~\bibnamefont {Thaler}},\ }\href {\doibase 10.1007/JHEP01(2019)121} {\bibfield  {journal} {\bibinfo  {journal} {JHEP}\ }\textbf {\bibinfo {volume} {01}},\ \bibinfo {pages} {121} (\bibinfo {year} {2019}{\natexlab{a}})},\ \Eprint {http://arxiv.org/abs/1810.05165} {arXiv:1810.05165 [hep-ph]} \BibitemShut {NoStop}%
\bibitem [{\citenamefont {Zaheer}\ \emph {et~al.}(2017)\citenamefont {Zaheer}, \citenamefont {Kottur}, \citenamefont {Ravanbakhsh}, \citenamefont {Poczos}, \citenamefont {Salakhutdinov},\ and\ \citenamefont {Smola}}]{Zaheer:2017wmg}%
  \BibitemOpen
  \bibfield  {author} {\bibinfo {author} {\bibfnamefont {M.}~\bibnamefont {Zaheer}}, \bibinfo {author} {\bibfnamefont {S.}~\bibnamefont {Kottur}}, \bibinfo {author} {\bibfnamefont {S.}~\bibnamefont {Ravanbakhsh}}, \bibinfo {author} {\bibfnamefont {B.}~\bibnamefont {Poczos}}, \bibinfo {author} {\bibfnamefont {R.}~\bibnamefont {Salakhutdinov}}, \ and\ \bibinfo {author} {\bibfnamefont {A.}~\bibnamefont {Smola}},\ }\href@noop {} {\  (\bibinfo {year} {2017})},\ \Eprint {http://arxiv.org/abs/1703.06114} {arXiv:1703.06114 [cs.LG]} \BibitemShut {NoStop}%
\bibitem [{\citenamefont {Qu}\ and\ \citenamefont {Gouskos}(2020)}]{Qu:2019gqs}%
  \BibitemOpen
  \bibfield  {author} {\bibinfo {author} {\bibfnamefont {H.}~\bibnamefont {Qu}}\ and\ \bibinfo {author} {\bibfnamefont {L.}~\bibnamefont {Gouskos}},\ }\href {\doibase 10.1103/PhysRevD.101.056019} {\bibfield  {journal} {\bibinfo  {journal} {Phys. Rev. D}\ }\textbf {\bibinfo {volume} {101}},\ \bibinfo {pages} {056019} (\bibinfo {year} {2020})},\ \Eprint {http://arxiv.org/abs/1902.08570} {arXiv:1902.08570 [hep-ph]} \BibitemShut {NoStop}%
\bibitem [{\citenamefont {Wang}\ \emph {et~al.}(2018)\citenamefont {Wang}, \citenamefont {Sun}, \citenamefont {Liu}, \citenamefont {Sarma}, \citenamefont {Bronstein},\ and\ \citenamefont {Solomon}}]{Wang:2018nkf}%
  \BibitemOpen
  \bibfield  {author} {\bibinfo {author} {\bibfnamefont {Y.}~\bibnamefont {Wang}}, \bibinfo {author} {\bibfnamefont {Y.}~\bibnamefont {Sun}}, \bibinfo {author} {\bibfnamefont {Z.}~\bibnamefont {Liu}}, \bibinfo {author} {\bibfnamefont {S.~E.}\ \bibnamefont {Sarma}}, \bibinfo {author} {\bibfnamefont {M.~M.}\ \bibnamefont {Bronstein}}, \ and\ \bibinfo {author} {\bibfnamefont {J.~M.}\ \bibnamefont {Solomon}},\ }\href@noop {} {\  (\bibinfo {year} {2018})},\ \Eprint {http://arxiv.org/abs/1801.07829} {arXiv:1801.07829 [cs.CV]} \BibitemShut {NoStop}%
\bibitem [{\citenamefont {Qu}\ \emph {et~al.}(2022)\citenamefont {Qu}, \citenamefont {Li},\ and\ \citenamefont {Qian}}]{Qu:2022mxj}%
  \BibitemOpen
  \bibfield  {author} {\bibinfo {author} {\bibfnamefont {H.}~\bibnamefont {Qu}}, \bibinfo {author} {\bibfnamefont {C.}~\bibnamefont {Li}}, \ and\ \bibinfo {author} {\bibfnamefont {S.}~\bibnamefont {Qian}},\ }\href {https://arxiv.org/abs/2202.03772} {\bibfield  {journal} {\bibinfo  {journal} {Proceedings of the 39th International Conference on Machine Learning}\ ,\ \bibinfo {pages} {18281}} (\bibinfo {year} {2022})},\ \Eprint {http://arxiv.org/abs/2202.03772} {arXiv:2202.03772 [hep-ph]} \BibitemShut {NoStop}%
\bibitem [{\citenamefont {Vaswani}\ \emph {et~al.}(2017)\citenamefont {Vaswani}, \citenamefont {Shazeer}, \citenamefont {Parmar}, \citenamefont {Uszkoreit}, \citenamefont {Jones}, \citenamefont {Gomez}, \citenamefont {Kaiser},\ and\ \citenamefont {Polosukhin}}]{Vaswani:2017lxt}%
  \BibitemOpen
  \bibfield  {author} {\bibinfo {author} {\bibfnamefont {A.}~\bibnamefont {Vaswani}}, \bibinfo {author} {\bibfnamefont {N.}~\bibnamefont {Shazeer}}, \bibinfo {author} {\bibfnamefont {N.}~\bibnamefont {Parmar}}, \bibinfo {author} {\bibfnamefont {J.}~\bibnamefont {Uszkoreit}}, \bibinfo {author} {\bibfnamefont {L.}~\bibnamefont {Jones}}, \bibinfo {author} {\bibfnamefont {A.~N.}\ \bibnamefont {Gomez}}, \bibinfo {author} {\bibfnamefont {L.}~\bibnamefont {Kaiser}}, \ and\ \bibinfo {author} {\bibfnamefont {I.}~\bibnamefont {Polosukhin}},\ }\href@noop {} {\bibfield  {journal} {\bibinfo  {journal} {31st International Conference on Neural Information Processing Systems}\ } (\bibinfo {year} {2017})},\ \Eprint {http://arxiv.org/abs/1706.03762} {arXiv:1706.03762 [cs.CL]} \BibitemShut {NoStop}%
\bibitem [{\citenamefont {Touvron}\ \emph {et~al.}(2021)\citenamefont {Touvron}, \citenamefont {Cord}, \citenamefont {Sablayrolles}, \citenamefont {Synnaeve},\ and\ \citenamefont {Jégou}}]{cord2021going}%
  \BibitemOpen
  \bibfield  {author} {\bibinfo {author} {\bibfnamefont {H.}~\bibnamefont {Touvron}}, \bibinfo {author} {\bibfnamefont {M.}~\bibnamefont {Cord}}, \bibinfo {author} {\bibfnamefont {A.}~\bibnamefont {Sablayrolles}}, \bibinfo {author} {\bibfnamefont {G.}~\bibnamefont {Synnaeve}}, \ and\ \bibinfo {author} {\bibfnamefont {H.}~\bibnamefont {Jégou}},\ }\href {https://arxiv.org/abs/2103.17239} {\bibfield  {journal} {\bibinfo  {journal} {IEEE/CVF International Conference on Computer Vision (ICCV)}\ ,\ \bibinfo {pages} {32}} (\bibinfo {year} {2021})},\ \Eprint {http://arxiv.org/abs/2103.17239} {arXiv:2103.17239 [cs.CV]} \BibitemShut {NoStop}%
\bibitem [{\citenamefont {Shleifer}\ \emph {et~al.}(2021)\citenamefont {Shleifer}, \citenamefont {Weston},\ and\ \citenamefont {Ott}}]{shleifer2021normformer}%
  \BibitemOpen
  \bibfield  {author} {\bibinfo {author} {\bibfnamefont {S.}~\bibnamefont {Shleifer}}, \bibinfo {author} {\bibfnamefont {J.}~\bibnamefont {Weston}}, \ and\ \bibinfo {author} {\bibfnamefont {M.}~\bibnamefont {Ott}},\ }\href {https://arxiv.org/abs/2110.09456} {\  (\bibinfo {year} {2021})},\ \Eprint {http://arxiv.org/abs/2110.09456} {arXiv:2110.09456 [cs.CL]} \BibitemShut {NoStop}%
\bibitem [{\citenamefont {Paszke}\ \emph {et~al.}(2019)\citenamefont {Paszke}, \citenamefont {Gross}, \citenamefont {Massa}, \citenamefont {Lerer}, \citenamefont {Bradbury}, \citenamefont {Chanan}, \citenamefont {Killeen}, \citenamefont {Lin}, \citenamefont {Gimelshein}, \citenamefont {Antiga} \emph {et~al.}}]{paszke2019pytorch}%
  \BibitemOpen
  \bibfield  {author} {\bibinfo {author} {\bibfnamefont {A.}~\bibnamefont {Paszke}}, \bibinfo {author} {\bibfnamefont {S.}~\bibnamefont {Gross}}, \bibinfo {author} {\bibfnamefont {F.}~\bibnamefont {Massa}}, \bibinfo {author} {\bibfnamefont {A.}~\bibnamefont {Lerer}}, \bibinfo {author} {\bibfnamefont {J.}~\bibnamefont {Bradbury}}, \bibinfo {author} {\bibfnamefont {G.}~\bibnamefont {Chanan}}, \bibinfo {author} {\bibfnamefont {T.}~\bibnamefont {Killeen}}, \bibinfo {author} {\bibfnamefont {Z.}~\bibnamefont {Lin}}, \bibinfo {author} {\bibfnamefont {N.}~\bibnamefont {Gimelshein}}, \bibinfo {author} {\bibfnamefont {L.}~\bibnamefont {Antiga}},  \emph {et~al.},\ }\href {https://arxiv.org/abs/1912.01703} {\bibfield  {journal} {\bibinfo  {journal} {Advances in neural information processing systems}\ }\textbf {\bibinfo {volume} {32}} (\bibinfo {year} {2019})},\ \Eprint {http://arxiv.org/abs/1912.01703} {arXiv:1912.01703 [cs.LG]} \BibitemShut {NoStop}%
\bibitem [{\citenamefont {Komiske}\ \emph {et~al.}(2019{\natexlab{b}})\citenamefont {Komiske}, \citenamefont {Metodiev},\ and\ \citenamefont {Thaler}}]{komiske2019energy}%
  \BibitemOpen
  \bibfield  {author} {\bibinfo {author} {\bibfnamefont {P.~T.}\ \bibnamefont {Komiske}}, \bibinfo {author} {\bibfnamefont {E.~M.}\ \bibnamefont {Metodiev}}, \ and\ \bibinfo {author} {\bibfnamefont {J.}~\bibnamefont {Thaler}},\ }\href {\doibase 10.1007/jhep01(2019)121} {\bibfield  {journal} {\bibinfo  {journal} {Journal of High Energy Physics}\ }\textbf {\bibinfo {volume} {2019}},\ \bibinfo {pages} {1} (\bibinfo {year} {2019}{\natexlab{b}})}\BibitemShut {NoStop}%
\bibitem [{\citenamefont {Mikuni}\ and\ \citenamefont {Canelli}(2021)}]{Mikuni:2021pou}%
  \BibitemOpen
  \bibfield  {author} {\bibinfo {author} {\bibfnamefont {V.}~\bibnamefont {Mikuni}}\ and\ \bibinfo {author} {\bibfnamefont {F.}~\bibnamefont {Canelli}},\ }\href {\doibase 10.1088/2632-2153/ac07f6} {\bibfield  {journal} {\bibinfo  {journal} {Mach. Learn. Sci. Tech.}\ }\textbf {\bibinfo {volume} {2}},\ \bibinfo {pages} {035027} (\bibinfo {year} {2021})},\ \Eprint {http://arxiv.org/abs/2102.05073} {arXiv:2102.05073 [physics.data-an]} \BibitemShut {NoStop}%
\bibitem [{\citenamefont {Ruhe}\ \emph {et~al.}(2023)\citenamefont {Ruhe}, \citenamefont {Brandstetter},\ and\ \citenamefont {Forr\'e}}]{Ruhe:2023rqc}%
  \BibitemOpen
  \bibfield  {author} {\bibinfo {author} {\bibfnamefont {D.}~\bibnamefont {Ruhe}}, \bibinfo {author} {\bibfnamefont {J.}~\bibnamefont {Brandstetter}}, \ and\ \bibinfo {author} {\bibfnamefont {P.}~\bibnamefont {Forr\'e}},\ }\href@noop {} {\  (\bibinfo {year} {2023})},\ \Eprint {http://arxiv.org/abs/2305.11141} {arXiv:2305.11141 [cs.LG]} \BibitemShut {NoStop}%
\bibitem [{\citenamefont {Bogatskiy}\ \emph {et~al.}(2022)\citenamefont {Bogatskiy}, \citenamefont {Hoffman}, \citenamefont {Miller},\ and\ \citenamefont {Offermann}}]{Bogatskiy:2022czk}%
  \BibitemOpen
  \bibfield  {author} {\bibinfo {author} {\bibfnamefont {A.}~\bibnamefont {Bogatskiy}}, \bibinfo {author} {\bibfnamefont {T.}~\bibnamefont {Hoffman}}, \bibinfo {author} {\bibfnamefont {D.~W.}\ \bibnamefont {Miller}}, \ and\ \bibinfo {author} {\bibfnamefont {J.~T.}\ \bibnamefont {Offermann}},\ }\href@noop {} {\  (\bibinfo {year} {2022})},\ \Eprint {http://arxiv.org/abs/2211.00454} {arXiv:2211.00454 [hep-ph]} \BibitemShut {NoStop}%
\bibitem [{\citenamefont {Spinner}\ \emph {et~al.}(2024)\citenamefont {Spinner}, \citenamefont {Bres\'o}, \citenamefont {de~Haan}, \citenamefont {Plehn}, \citenamefont {Thaler},\ and\ \citenamefont {Brehmer}}]{Spinner:2024hjm}%
  \BibitemOpen
  \bibfield  {author} {\bibinfo {author} {\bibfnamefont {J.}~\bibnamefont {Spinner}}, \bibinfo {author} {\bibfnamefont {V.}~\bibnamefont {Bres\'o}}, \bibinfo {author} {\bibfnamefont {P.}~\bibnamefont {de~Haan}}, \bibinfo {author} {\bibfnamefont {T.}~\bibnamefont {Plehn}}, \bibinfo {author} {\bibfnamefont {J.}~\bibnamefont {Thaler}}, \ and\ \bibinfo {author} {\bibfnamefont {J.}~\bibnamefont {Brehmer}},\ }\href@noop {} {\  (\bibinfo {year} {2024})},\ \Eprint {http://arxiv.org/abs/2405.14806} {arXiv:2405.14806 [physics.data-an]} \BibitemShut {NoStop}%
\bibitem [{\citenamefont {Gong}\ \emph {et~al.}(2022)\citenamefont {Gong}, \citenamefont {Meng}, \citenamefont {Zhang}, \citenamefont {Qu}, \citenamefont {Li}, \citenamefont {Qian}, \citenamefont {Du}, \citenamefont {Ma},\ and\ \citenamefont {Liu}}]{Gong:2022lye}%
  \BibitemOpen
  \bibfield  {author} {\bibinfo {author} {\bibfnamefont {S.}~\bibnamefont {Gong}}, \bibinfo {author} {\bibfnamefont {Q.}~\bibnamefont {Meng}}, \bibinfo {author} {\bibfnamefont {J.}~\bibnamefont {Zhang}}, \bibinfo {author} {\bibfnamefont {H.}~\bibnamefont {Qu}}, \bibinfo {author} {\bibfnamefont {C.}~\bibnamefont {Li}}, \bibinfo {author} {\bibfnamefont {S.}~\bibnamefont {Qian}}, \bibinfo {author} {\bibfnamefont {W.}~\bibnamefont {Du}}, \bibinfo {author} {\bibfnamefont {Z.-M.}\ \bibnamefont {Ma}}, \ and\ \bibinfo {author} {\bibfnamefont {T.-Y.}\ \bibnamefont {Liu}},\ }\href {\doibase 10.1007/JHEP07(2022)030} {\bibfield  {journal} {\bibinfo  {journal} {JHEP}\ }\textbf {\bibinfo {volume} {07}},\ \bibinfo {pages} {030} (\bibinfo {year} {2022})},\ \Eprint {http://arxiv.org/abs/2201.08187} {arXiv:2201.08187 [hep-ph]} \BibitemShut {NoStop}%
\bibitem [{\citenamefont {Mikuni}\ and\ \citenamefont {Canelli}(2020)}]{Mikuni:2020wpr}%
  \BibitemOpen
  \bibfield  {author} {\bibinfo {author} {\bibfnamefont {V.}~\bibnamefont {Mikuni}}\ and\ \bibinfo {author} {\bibfnamefont {F.}~\bibnamefont {Canelli}},\ }\href {\doibase 10.1140/epjp/s13360-020-00497-3} {\bibfield  {journal} {\bibinfo  {journal} {Eur. Phys. J. Plus}\ }\textbf {\bibinfo {volume} {135}},\ \bibinfo {pages} {463} (\bibinfo {year} {2020})},\ \Eprint {http://arxiv.org/abs/2001.05311} {arXiv:2001.05311 [physics.data-an]} \BibitemShut {NoStop}%
\bibitem [{\citenamefont {Dreyer}\ and\ \citenamefont {Qu}(2021)}]{Dreyer:2020brq}%
  \BibitemOpen
  \bibfield  {author} {\bibinfo {author} {\bibfnamefont {F.~A.}\ \bibnamefont {Dreyer}}\ and\ \bibinfo {author} {\bibfnamefont {H.}~\bibnamefont {Qu}},\ }\href {\doibase 10.1007/JHEP03(2021)052} {\bibfield  {journal} {\bibinfo  {journal} {JHEP}\ }\textbf {\bibinfo {volume} {03}},\ \bibinfo {pages} {052} (\bibinfo {year} {2021})},\ \Eprint {http://arxiv.org/abs/2012.08526} {arXiv:2012.08526 [hep-ph]} \BibitemShut {NoStop}%
\bibitem [{\citenamefont {Tumasyan}\ \emph {et~al.}(2023)\citenamefont {Tumasyan} \emph {et~al.}}]{CMS:2022psv}%
  \BibitemOpen
  \bibfield  {author} {\bibinfo {author} {\bibfnamefont {A.}~\bibnamefont {Tumasyan}} \emph {et~al.} (\bibinfo {collaboration} {CMS}),\ }\href {\doibase 10.1103/PhysRevLett.131.061801} {\bibfield  {journal} {\bibinfo  {journal} {Phys. Rev. Lett.}\ }\textbf {\bibinfo {volume} {131}},\ \bibinfo {pages} {061801} (\bibinfo {year} {2023})},\ \Eprint {http://arxiv.org/abs/2205.05550} {arXiv:2205.05550 [hep-ex]} \BibitemShut {NoStop}%
\end{thebibliography}%

\end{document}